\documentstyle[pra,aps,psfig,epsfig]{revtex}

\begin{document}

\title{Adiabatic Output Coupling of a Bose Gas at Finite Temperatures}
\author{S. Choi, Y. Japha, and  K. Burnett}
\address{Clarendon Laboratory,
Department of Physics, University of Oxford, Parks Road,
\mbox{Oxford OX1 3PU,  United Kingdom.}}

\vspace{6mm}
\maketitle

\begin{abstract}
We develop a general theory of adiabatic output coupling from trapped,  
weakly-interacting, atomic
Bose-Einstein Condensates at finite temperatures. For weak coupling, the
output rate from the condensate and the excited levels in the trap settles
in  a time proportional to the inverse of the spectral width of
the coupling to the output modes. We discuss the properties of the output
atoms in the quasi-steady-state where the population inside the trap is not
appreciably
depleted. We show how the composition of the output
beam, containing the condensate and the thermal component, may be
controlled by changing the frequency of the output coupling lasers.
This composition determines the first and second order coherence of the
output beam. We discuss the changes in the composition of the Bose gas
left in the trap and show how non-resonant output coupling can stimulate
either the evaporation of the thermal excitations in the trap or the growth 
of
the non-thermal excitations, when pairs of correlated atoms leave the
condensate.
\end{abstract}

\section{Introduction}

Trapped atomic Bose-Einstein Condensates (BEC) are now routinely
produced in various laboratories around the world, and it is important to
understand the factors that influence the coherence of atoms
transferred from them. This is an essential issue for the atom laser
research, which has the long-term goal of producing continuous,
directional, and coherent beams of atoms.  A matter-wave pulse was
first produced by using a radio frequency (RF) electromagnetic pulse
to transfer atoms out of a
trap, where they were
allowed to fall freely under 
gravity\cite{MITOutput97,MITfringes97,NewZealand99}.  More recently, a 
stimulated Raman process induced a
transition to an untrapped magnetic state in an
experiment at NIST; net momentum kick
provided by the process resulted in a highly directional beam\cite{NIST99}.
On the other hand, a long beam of atoms falling under gravity was produced 
in Munich using an RF-field-induced transition\cite{Munich99}.
It is noted that although the more general features of the output in these 
experiments are fairly well-understood, detailed properties of the atoms in 
the output beam, and the evolution of the component that remains inside the 
trap have not so far been investigated.

Previous theoretical treatments of the output couplers for condensates
have been either limited to a single-mode non-interacting trapped
condensate\cite{Hop97,MoySav97,JefHorBar99} or to mean field treatment
for the
condensate\cite{BalBurSco97,Steck98,Durham98,Jack99,Band99,Edwards99} which
assume that the output beam is extracted out of a condensate at zero
temperature and that they can be described by a single complex function of 
space
and time.  However, real condensates appear at finite temperatures,
and as a result, thermal excitations play a major role.

In a previous paper\cite{JapChoiBur99}, we outlined a theory of weak
output coupling from a partially condensed, trapped Bose gas at finite
temperatures. By applying the self-consistent Hartree-Fock-Bogoliubov
(HFB) theory for Bose gases at finite temperatures, we identified
three output components. The first is the
output of pure condensates, which we have called ``coherent output.'' The
second is the fraction emerging from
the thermal excitations in the trap, which are coupled out of the trap by
the process of ``stimulated quantum evaporation.''  This is equivalent to 
the quantum evaporation of Helium atoms from the surface of
superfluid $^4$He, where phonon excitations travel up to the
surface of the superfluid and then spontaneously emerge from the surface
as evaporated atoms\cite{Qevap}. In our case such an evaporation from
the trap is stimulated by an electromagnetic field. The last
output component comes from the
process of ``pair breaking,'' which involves simultaneous creation of an
output coupled atom and an elementary excitation (quasi-particle) within the
trap. For suitable choices of the coupling parameters each of the three
processes can become the dominant process.
We have shown that output coupling can serve not only as a
useful way to extract an atomic beam out of a trap, but also as a
probe to the delicate features of the quantum state of the Bose gas,
including the pair correlations inside the condensate.

In this paper, we present an extensive analysis of the spectrum of the
output atoms, and address issues that were not included in our shorter
work. The first is the conditions for the output coupling to give a
steady flow of atoms. We discuss the behaviour of the output rate and
the atomic density in the short and the long time regimes, and also discuss 
the conditions
for achieving a steady output beam. Second, the application of output
coupling must cause changes in the state of the trapped Bose gas, such
as changes of the number of excitations relative to the number of
condensate atoms in the trap. We present a thorough discussion of
these changes.  The state of the Bose gas in the trap is usually
described by the Bogoliubov formalism, which assumes an indefinite
number of atoms in the system and therefore is not number-conserving. In 
this paper
we discuss a number-conserving description of the system, which is
especially useful when we consider the process of pair-breaking.

The results of this paper are directly applicable for any output
coupling scheme which involves a single trapped state and one output state.
We demonstrate here general fundamental
issues by considering a one-dimensional Bose gas in a harmonic
potential, which is coupled into a free output level in the absence of
gravity.

The structure of this paper is as follows: We begin by deriving the 
equations of motion for the evolution of dynamical variables inside and 
outside the trap in
Section~\ref{sec:gen}. We present in
Section~\ref{sec:out} a quasi-steady-state formalism, which
enables us to obtain the properties of the output atoms. We
demonstrate the results by a numerical example. We
outline in Section~\ref{sec:dynamics} the solution to the equations of
motion in the trap that were derived in Section~\ref{sec:gen}, by 
introducing a number-conserving, time dependent Hartree-Fock-Bogoliubov 
(HFB) formulation
in an adiabatic approximation. Applying this, we obtain expressions
for the internal modes of the system in two different regimes, from which 
the time dependent quasiparticle excitations can be calculated. Finally,
discussions and summary are given in Section V.

\section{The two-state output coupling model}
\label{sec:gen}

In this section we present our model for describing the output coupling of a 
trapped Bose gas into free output modes. We derive the equations of motion
for the atomic field operators in the trapped and the untrapped states, and 
give a general form of their solutions.

\subsection{Description of the model}

Our model assumes that atoms  are initially in an atomic magnetic level $| t \rangle$ (``the trapped state'') confined by a potential and in  
thermal equilibrium.  A coupling interaction is then switched on, inducing
transitions to a different magnetic level $| f \rangle$ (the ``free'' or 
``untrapped''
state). We stress that these are labels denoting internal atomic levels, not 
the centre of mass states, so that for short enough times an atom in an $| f \rangle$ state may still be 
present within the trap.  We use the atomic field
operator $\hat{\psi}_t({\bf r})$ to describe the amplitude for the
annihilation of a trapped atom at point ${\bf r}$, and the 
operator $\hat{\psi}_f({\bf r})$ to describe the corresponding amplitude for
a free, untrapped atom. The Hamiltonian of the system takes the form
\begin{equation}
\hat{{\cal H}}=\hat{{\cal H}}_t^{(0)} +\hat{{\cal H}}_f^{(0)}
+ \hat{{\cal H}}_{\rm couple},
\label{Ham}
\end{equation}
where $\hat{{\cal H}}_t^{(0)}$ and $\hat{{\cal H}}_f^{(0)}$ describe the
dynamics of the trapped and untrapped atoms respectively while $\hat{{\cal
H}}_{\rm couple}$ describes the coupling between the two states.

The dynamics inside the trap are given by the many-body Hamiltonian:
\begin{eqnarray}
\hat{{\cal H}}_t^{(0)} & = &
\int d^3 {\bf  r} \hat{\psi}_t^{\dag}({\bf 
r})\left[-\frac{\hbar^2\nabla^2}{2m}
+V_t({\bf  r})\right] \hat{\psi}_t({\bf  r}) \nonumber \\
&&+\frac{1}{2}\int d^3 {\bf  r} \int d^3 {\bf  r}' \hat{\psi}_t^{\dag}({\bf
r})\hat{\psi}_t^{\dag}({\bf  r}')
U_{tt}({\bf  r}-{\bf  r}')\hat{\psi}_t({\bf  r}')\hat{\psi}_t({\bf  r}),
\end{eqnarray}
where $m$ is the mass of a single atom, $V_t({\bf r})$ is the
potential responsible for the confinement of the atoms in the trap and
$U_{tt}$ is the inter-particle potential between the trapped
atoms.

With the output atoms, significant effect of their (elastic)
collisions with the trapped atoms must be taken into account.  In
addition, we consider a small rate of output from the trap, so that
the output atoms are dilute; this enables us to neglect the
interactions between the free atoms themselves. Since the Bose gases
are typically so dilute that their mean-free-path for the inelastic
collisions is much larger than the dimensions of the atomic cloud in the
trap, one may neglect also any inelastic collisions of the output
atoms with the trapped atoms\cite{Wilkens-scatt}.
The Hamiltonian for the output atoms is
then given by
\begin{eqnarray}
\hat{{\cal H}}_f^{(0)} & = & \int d^3 {\bf  r} \hat{\psi}_f^{\dag}({\bf
r})\left[-\frac{\hbar^2\nabla^2}{2m}
+V_f({\bf  r})\right] \hat{\psi}_f({\bf  r}) \nonumber \\
&& +\int d^3 {\bf  r} \int d^3 {\bf  r}' \hat{\psi}_f^{\dag}({\bf
r}) \hat{\psi}_t^{\dag}({\bf  r}')
U_{tf}({\bf  r}-{\bf  r}') \hat{\psi}_f({\bf  r}') \hat{\psi}_t({\bf  r}),
\end{eqnarray}
where $V_f({\bf r})$ is the potential that influences the propagation
of the output free atoms and $U_{tf}$ is the collisional interaction
between the trapped and free atoms; this is in general different from
the interaction $U_{tt}$.  We use the usual $\delta$-function form for the
inter-particle potentials
\begin{eqnarray}
U_{tt}({\bf  r}-{\bf  r}') & = & U_{0}\delta({\bf  r}-{\bf  r}') \\
U_{tf}({\bf  r}-{\bf  r}') & = & U_{1}\delta({\bf  r}-{\bf  r}'),
\end{eqnarray}
where $U_0=4\pi \hbar^2 a_{tt}/m$ and $U_{1}=4\pi \hbar^2 a_{tf}/m$ are
proportional to the $s$-wave scattering lengths $a_{tt}$ and $a_{tf}$ for
trapped-trapped and trapped-free collisions, respectively.
We assume a repulsive interaction between the atoms, i.e. $U_{0,1} > 0$.

\begin{figure}[t] \begin{center}
\epsfig{width=9cm,file=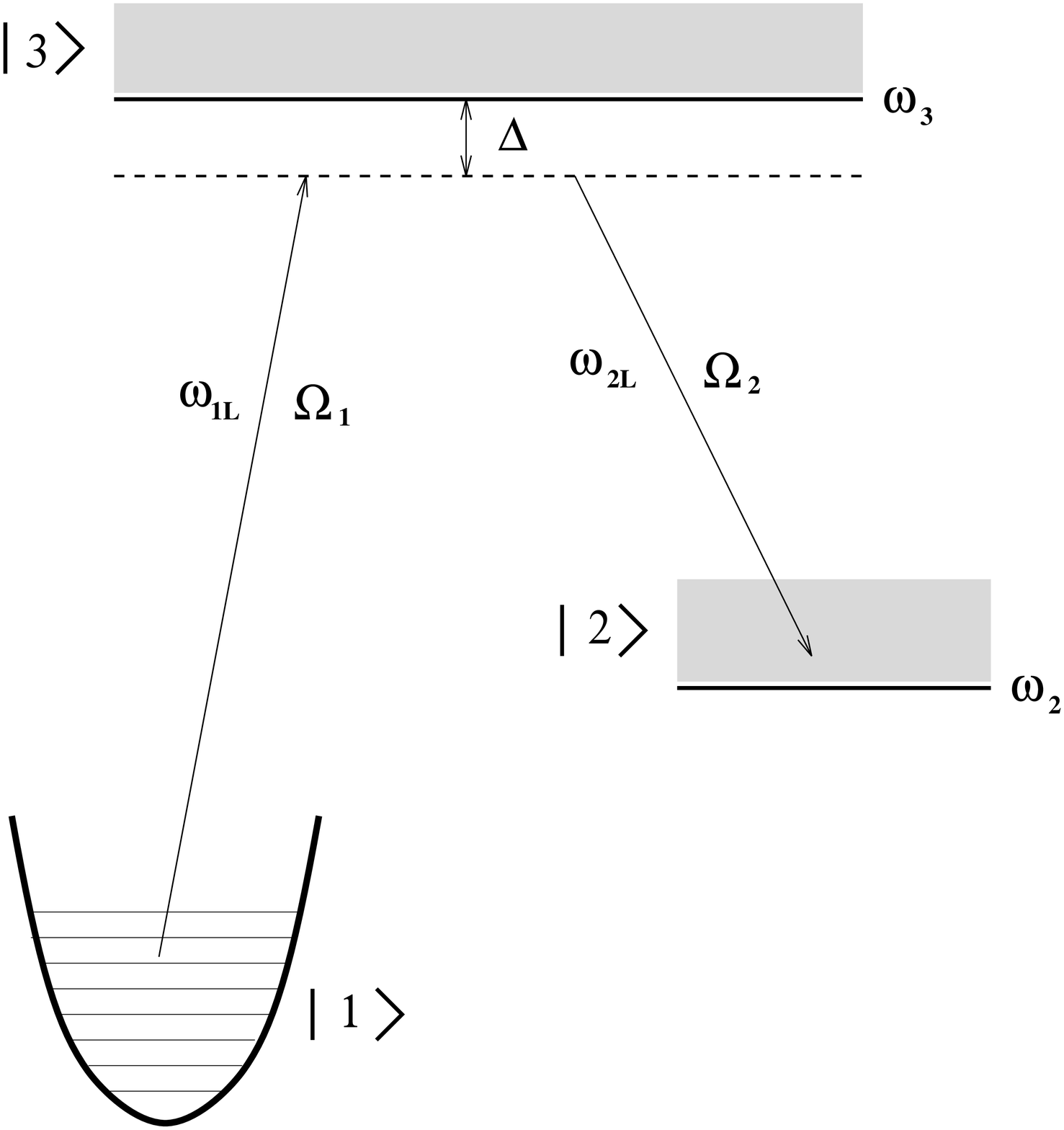}
\end{center}
\caption{\protect \footnotesize  Schematic diagram of energy levels and
couplings involved in the stimulated Raman process. $\Delta$ is the 
detuning,
$\omega_{i}$, $\Omega_{iL}$, $i = 1,2$ are the frequencies and the Rabi 
frequencies of the two lasers.}
\label{raman}
\end{figure}

For the Hamiltonian $\hat{{\cal H}}_{\rm couple}$, we consider
coupling by an electromagnetic (EM) field which induces transitions
between the states $| t \rangle$ and $| f \rangle$.  In the rotating
wave approximation, the EM coupling mechanism may be described by the
following Hamiltonian, which can quite clearly be generalised to describe 
any
kind of linear coupling such as weak tunnelling:
\begin{equation}
\hat{{\cal H}}_{\rm couple} =  \hbar\int d^3 {\bf  r} \lambda({\bf
r},t) \hat{\psi}_f^{\dag}({\bf  r}) \hat{\psi}_t({\bf  r}) + h.c.
\end{equation}
Here $\lambda({\bf r},t)$ denotes the amplitude of coupling between
the trapped and the untrapped magnetic states. The form of $\lambda({\bf
r},t)$ depends on the type of coupling used: Typical mechanisms are
direct (one-photon) radio-frequency transition and indirect
(two-photon) stimulated Raman transition.  For any EM induced processes,
the coupling can be written as
\begin{equation}
\lambda({\bf  r},t)=\bar{\lambda}({\bf  r},t)e^{i({\bf k}_{\rm em}\cdot{\bf 
r}
- \Delta_{em} t)},
\end{equation}
where $\bar{\lambda}$ is slowly varying in space and
time. $\bar{\lambda}$ can be either time-independent, to describe a
continuous electromagnetic wave, or pulsed. $\hbar{\bf k}_{\rm
em}$ and $\hbar\Delta_{\rm em}$ measure the net momentum and energy
transfer from the EM field to an output atom.  In an RF coupling
scheme, $\bar{\lambda}({\bf  r},t)$ is the Rabi frequency $\Omega({\bf r},t) 
=\langle
\hat{p}\rangle {\cal E}({\bf r},t)/\hbar$ (or $\langle
\hat{\mu}\rangle {\cal B}({\bf r},t)/\hbar$) corresponding to the
flipping of the atomic electric (or magnetic) dipole $\langle
\hat{p}\rangle$ (or $\langle \hat{\mu}\rangle $) in the electric (or
magnetic) field ${\cal E}({\bf r},t)$ (${\cal B}({\bf r},t)$);
$\Delta_{\rm em}$ is the detuning of the EM field frequency from the
transition frequency and ${\bf k}_{\rm em}$ is, in general, negligible 
compared to the initial momentum distribution of the atoms.  In the 
stimulated
Raman coupling, two laser beams are used to induce a transition from
$|t \rangle$ to $| f \rangle$ through an intermediate level $| i
\rangle$, and
\begin{equation}
\bar{\lambda}({\bf  r},t)=\frac{\Omega^*_{ti}({\bf  r},t)\Omega_{fi}({\bf
r},t)}{\Delta_{i}} ,
\label{lambda}
\end{equation}
where $\Omega_{ti}$ and $\Omega_{fi}$ are the Rabi frequencies
corresponding to the intermediate transitions and $\Delta_i$ is their
detuning from resonance with the two beams.  $\Delta_{\rm em}$ and
${\bf k}_{\rm em}$ are the differences between the frequencies and
momenta associated with the two laser beams:
\begin{eqnarray}
\Delta_{em} & = & \omega_{1L} - \omega_{2L}  -  \frac{E_t^{(0)} - E_f^{(0)}}
{\hbar},\\
{\bf k}_{em} & = & {\bf k}_{1L} - {\bf k}_{2L} ,
\end{eqnarray}
where $E_t^{(0)} - E_f^{(0)}$
is the energy splitting between the atomic levels
$| t \rangle$ and $| f \rangle$ in the centre of the trap.
A more detailed derivation of Eq.~(\ref{lambda})
for the Raman process is provided in Appendix~\ref{app:Raman}.
An energy level diagram depicting the output coupling through the stimulated
Raman process is given in Fig.~\ref{raman}.

\subsection{Equations of motion}

The coupled equations of motion for the trapped and free field
operators are obtained by computing their commutation relations with
the Hamiltonian~(\ref{Ham}).  We thus find
\begin{eqnarray}
\frac{\partial}{\partial t} \hat{\psi}_t({\bf r}, t)&=& 
-\frac{i}{\hbar}{\cal
L}_t
\hat{\psi}_t({\bf r},t)
-\frac{i}{\hbar}U_0\hat{\psi}_t^{\dag}({\bf r},t)\hat{\psi}_t({\bf r},t)
\hat{\psi}_t({\bf r},t) \nonumber \\ &&
-i\lambda^*({\bf r},t)\hat{\psi}_f({\bf r},t),
\label{eq:psit} \\
\frac{\partial}{\partial t}{\hat{\psi}}_f({\bf r},t)&=& 
-\frac{i}{\hbar}{\cal
L}_f
\hat{\psi}_f({\bf r},t)
-i\lambda({\bf r},t)\hat{\psi}_t({\bf r},t),
\label{eq:psif}
\end{eqnarray}
where
\begin{eqnarray} 
{\cal L}_t &\equiv& -\hbar^2\nabla^2/2m+V_t({\bf r}),
\label{calLt} \\
{\cal L}_f & \equiv &  -\hbar^2\nabla^2/2m+V_f({\bf r})
+ U_1\langle \hat{\psi}_t^{\dag}({\bf r})\hat{\psi}_t({\bf r})\rangle.\label{calLt2}
\end{eqnarray}
In Eq.~(\ref{calLt2}) we have used a mean-field approximation for the collisional
effect of the trapped atoms on the untrapped ones. This approximation
neglects inelastic scattering processes with the trapped
atoms\cite{Wilkens-scatt} and other possible effects of entanglement
of the output atoms with the atoms in the trap. The approximation is
justified under the assumption of long atomic mean-free-path mentioned above.

\subsection{General solutions}

\subsubsection{Output atoms}

The formal solution of Eq.~(\ref{eq:psif}) for $\hat{\psi}_f$ in terms
of $\hat{\psi}_t$ is
\begin{equation}
\hat{\psi}_f({\bf r},t) = \hat{\psi}_f^{(0)}({\bf r},t)
-i\int_0^t dt'\int d^3{\bf r}' K_f({\bf r},{\bf r}',t-t')\lambda({\bf 
r}',t')
\hat{\psi}_t({\bf r}',t'),
\label{solf}
\end{equation}
where $\hat{\psi}_f^{(0)}$
satisfies the time-dependent Schr\"odinger
equation for the free evolution of $\hat{\psi}_f$ in the absence of output
coupling
\begin{equation} \frac{\partial}{\partial
t}\hat{\psi}_f^{(0)}=-\frac{i}{\hbar}{\cal
L}_f\hat{\psi}_f^{(0)},
\label{psif0} \end{equation}
and the ``free'' propagator $K_f({\bf r},{\bf r}',t-t')$
satisfies the partial differential equation
\begin{equation} \frac{\partial}{\partial t} K_f({\bf  r},{\bf
r}',t-t')=-\frac{i}{\hbar}{\cal L}_f
K_f({\bf  r},{\bf  r}',t-t')+\delta({\bf  r}-{\bf  r}')\delta(t-t').
\end{equation}
The second term in Eq.~(\ref{solf}) describes the transition of the atoms
from the trapped level $| t \rangle$ into the free level $| f \rangle$
with amplitude $\lambda({\bf r},t)$ and their subsequent propagation
as free atoms.

We note that it is useful to expand the field operator $\hat{\psi}_f$ in 
terms of
the normal
modes $\varphi_{{\bf  k}}({\bf  r})$ of the untrapped level
\begin{equation}
\hat{\psi}_f({\bf  r}, t)=\sum_{{\bf  k}}\varphi_{{\bf  k}}({\bf  
r})\hat{b}_{{\bf  k}}(t),
\label{psif-expan}
\end{equation}
where $\hat{b}_{{\bf k}}$ satisfy the usual bosonic commutation relations
$[\hat{b}_{{\bf k}}, \hat{b}_{{\bf k}'}^{\dag}]=\delta_{{\bf k},{\bf k}'}$. 
${\bf k}$ denotes the momentum state of the free atoms with energy 
$E_{{\bf k}}=\hbar\omega_{{\bf k}}$, and $\varphi_{{\bf k}}$ are the
time-independent solutions of the single-particle problem in the
non-trapping effective potential $V_f({\bf r})+U_1 \langle
\hat{\psi}_t^{\dag}({\bf r})\hat{\psi}_t({\bf r}) \rangle$ created by the 
mean
field effect of the collisions between the trapped and the free
atoms. In principle, the solutions $\varphi_{{\bf k}}$ and the
energies $E_{{\bf k}}$ may change with time due to the change in the
density of the trapped atoms and the subsequent change in the
effective repulsive potential near the trap. In what follows we will
neglect this time-dependence under the assumption of weak output
coupling and slow changes in the density of the trapped atoms.

In the absence of gravitational or other forces, at positions far away
from the trap, ${\bf k}$ may be taken to be the wave number of a plane
wave $\varphi_{{\bf k}} \sim e^{i{\bf k}\cdot{\bf r}}$ with
$\omega_{{\bf k}}=\hbar k^2/2m$.  In the presence of gravity, the structure of the output modes should be defined appropriately, and 
the modes ${\bf k}$ may be given asymptotically by the solutions of
the Schr\"odinger equation in a homogeneous field.  In
Eq.~(\ref{psif-expan}) we have used a sum $\sum_{{\bf k}}$ over
discrete output states. It is noted that the actual structure of the Hilbert space for the
output modes depends on the potential $V_f({\bf r})$; if this
potential vanishes far away from the centre of the trap, then the sum
$\sum_{{\bf k}}$ should be replaced by an integral $\int d^3 {\bf k}$
and the operators $\hat{b}_{{\bf k}}$ should be defined such that 
$[\hat{b}_{{\bf
k}},\hat{b}^{\dag}_{{\bf k}'}]=\delta({\bf k}-{\bf k}')$.  

In terms of the basis functions $\varphi_{{\bf k}}({\bf
r})$, the free
field operator $\hat{\psi}_f^{(0)}$ is given by
\begin{equation}
\hat{\psi}_f^{(0)}({\bf  r},t)=\sum_{{\bf  k}} \varphi_{{\bf
k}}({\bf  r}) \hat{b}_{{\bf  k}}(0)
e^{-i\omega_{{\bf  k}}t} \label{psif0b}
\end{equation}
and the propagator of the free atoms may be written as
\begin{equation} 
K_f({\bf r},{\bf r}',t-t')=\sum_{{\bf k}} \varphi_{{\bf
k}}({\bf
r})\varphi^*_{{\bf k}}({\bf r}') e^{-i\omega_{{\bf k}}(t-t')}\theta(t-t') .
\label{Gfk}
\end{equation}
It is useful to describe the evolution of the untrapped atoms in terms
of the annihilation operators $\hat{b}_{{\bf  k}}$ in a specific mode ${\bf  k}$. The solution for this operator is obtained by
multiplying
Eq.~(\ref{solf}) by $\varphi^*_{{\bf  k}}$ and integrating over all space:
\begin{equation}
\hat{b}_{{\bf  k}}(t)= \hat{b}_{{\bf  k}}(0)e^{-i\omega_{{\bf  k}} t}
-i\int_0^t dt'\int d^3{\bf r}\varphi^*_{{\bf  k}}({\bf  r})\lambda({\bf
r},t')
e^{-i\omega_{{\bf  k}}(t-t')}
\hat{\psi}_t({\bf r},t').
\label{solbk}
\end{equation}

Solutions for the output field $\hat{\psi}_t({\bf r},t)$ in 
Eqs.~(\ref{solf}) and~(\ref{solbk})
require an explicit expression for the trapped field operator
$\hat{\psi}_t({\bf r},t')$. The simplest approximation is to take the first 
order solution in the coupling amplitude $\lambda({\bf r},t)$. This 
corresponds to a
very weak coupling and $\hat{\psi}_t({\bf r},t) \approx 
\hat{\psi}_t^{(0)}({\bf
r},t)$, where $\hat{\psi}_t^{(0)}({\bf r},t)$ is the field operator of the 
trapped atoms without output coupling.  The fundamental properties of the
output under such approximation will be the main subject of
Sec.~\ref{sec:out}.

\subsubsection{Trapped atoms}

By substituting the solution~(\ref{solf}) for $\hat{\psi}_f$ back into
Eq.~(\ref{eq:psit}) we obtain the following equation for the trapped
field operator $\hat{\psi}_t({\bf r},t)$
\begin{eqnarray}
\frac{\partial}{\partial t}\hat{\psi}_t({\bf  r},t) &=& 
-\frac{i}{\hbar}{\cal
L}_t
\hat{\psi}_t({\bf r},t)
-\frac{i}{\hbar}U_0\hat{\psi}_t^{\dag}({\bf r},t)\hat{\psi}_t({\bf r},t)
\hat{\psi}_t({\bf r},t)  \nonumber \\
&& -\int_0^t dt' \int d^3 {\bf  r} G({\bf  r},{\bf  r}',t,t')\psi_t({\bf
r}',t')
-i\lambda^*({\bf r},t)\hat{\psi}_f^{(0)}({\bf r},t),
\label{eq:psi0}
\end{eqnarray}
where
\begin{equation}
G({\bf  r},{\bf  r}',t,t')= \lambda^*({\bf  r},t)K_f({\bf  r},{\bf
r}',t-t')\lambda({\bf  r}',t').
\label{Gamma}
\end{equation}

The solution of the integro-differential equation, Eq.~(\ref{eq:psi0}) will
be the main subject of Sec.~\ref{sec:dynamics}. In principle, two
different situations may be expected from such an integro-differential
equation. In the case where this equation describes a coupling to the output
levels with a narrow available bandwidth compared to the coupling
strength $\lambda({\bf r},t)$, we anticipate Rabi oscillations of the atomic
population between the trapped and untrapped levels. Physically, this
means that the output atoms stay near the trap for a long enough time
to perform these oscillations.  On the other hand, when the bandwidth of the
output modes is large compared to the coupling strength, an
exponential decay of the population in the trap is
expected. Physically, this behaviour is expected when the output atoms
are fast enough to escape from the trap before the interaction couples
them back into the trapped level.  Even if the coupling is very weak,
an oscillatory kind of behaviour is expected for short times compared
to the inverse of the bandwidth of the relevant output modes, before
the output rate settles on a constant rate with fixed energy.  For this
last case, Eq. (\ref{eq:psi0}) may be viewed as a Langevin equation for
an interaction of a confined system with an infinite heat
bath\cite{Jack99}.

The trapped field operators $\hat{\psi}_t({\bf r},t)$ can, in principle, be expanded similarly
to the field operator $\hat{\psi}_f({\bf  r},t)$ as
\begin{equation}
\hat{\psi}_t({\bf  r},t)=\sum_n \phi_n({\bf  r})\hat{a}_n(t),
\end{equation}
where $\phi_n({\bf r})$ are the normal modes of the trap given by the
solutions of the single-particle problem in the potential well
$V_t({\bf r})$. These eigenmodes, however, do not form a good basis,
since the interaction between the atoms results in a strong mixing
between the levels.  In the following sections, therefore, we will instead 
use the basis
of condensate and its quasi-particle excitations, which are obtained from the
Hartree-Fock-Bogoliubov theory of an interacting Bose gas.

\section{Output properties in the quasi-steady state}
\label{sec:out}

In this section we present the properties of the output atoms under
the assumption that the output coupling is very weak. In this case,
the output beam of atoms can serve as a probe to the structure of the
Bose gas in the trap under steady-state conditions. The solution for
the output is given by Eq.~(\ref{solf}), or equivalently Eq.~(\ref{solbk}),
where we
substitute $\hat{\psi}_t({\bf r},t) \approx \hat{\psi}_t^{(0)}({\bf r},t)$.  
We start this section by giving a brief description of the formalism that allows
the use of the steady-state solution $\hat{\psi}_t^{(0)}({\bf r},t)$ for the
atoms in the trap. We then present basic properties of the output, in 
particular
the spectrum and the density of the output coupled atoms. Finally the first 
and second order
correlations of the output atoms are presented.

\subsection{The trapped atoms in the quasi steady-state}
\label{trap-steady}

We briefly review in this subsection the theory of the trapped Bose
gas in steady-state conditions, in a way that will enable us in
Sec.~\ref{sec:dynamics} to extend the theory to the time-dependent
case where the number of atoms in the trap does change adiabatically
during output coupling. Since the situations discussed in this paper involve 
transitions of atoms into untrapped propagating states where counting
the number of output atoms could be one of the main measurements,
we choose to use a number-conserving
theory in the spirit of the theories put forward recently\cite{numconserve}.
In addition, the theory enables an extension of the finite temperature 
HFB-Popov
method into time dependent cases.

For describing a partially condensed system of atoms at a finite
temperature, we split the field operator for the atoms in the trap
into a part which is proportional to the condensate
wave function and a part which represents excitations orthogonal to
this state:
\begin{equation}
\hat{\psi}_t({\bf  r},t)=e^{-i\Phi(t)}\left[\psi_0({\bf  
r})\hat{a}_0(t)+\delta\hat{\psi}({\bf
r},t)
\right]. \label{psit}
\end{equation}
Here $\Phi(t)$ is a global phase given by
\begin{equation}
\Phi(t)=\int^t_{0} \mu(t')dt' ,
\end{equation}
where $\mu$, the chemical potential of the system for the given global
variables, is constant under steady-state conditions. The operator
$\hat{a}_0$ is a bosonic annihilation operator satisfying
$[\hat{a}_0,\hat{a}_0^{\dag}]=1$ and describes the annihilation of one atom 
in the
condensate state $\psi_0({\bf  r})$. The number of condensate atoms is
represented by the operator $\hat{N}_0\equiv \hat{a}_0^{\dag}\hat{a}_0$.  
The
non-condensate part, $\delta \hat{\psi}$, is assumed to be orthogonal to the
condensate in the sense $\int d^3 {\bf r} \hat{\psi}^*_0({\bf
r},t)\delta\hat{\psi}({\bf r},t)=0$. In the number conserving formalism 
$\delta \hat{\psi}$
is approximated by the following Bogoliubov form:
\begin{equation}
\delta \hat{\psi}({\bf  r},t)\approx \hat{a}_0\frac{1}{\sqrt{\hat{N}_0}}
\sum_j[u_j({\bf  r})\hat{\alpha}_j(t)+v^*_j({\bf  
r})\hat{\alpha}^{\dag}_j(t)],
\label{dpsit}
\end{equation}
where $\hat{\alpha}_j,\hat{\alpha}^{\dag}_j$ are the bosonic operators 
satisfying
$[\hat{\alpha}_i, \hat{\alpha}_j^{\dag}]=\delta_{ij}$ in the space of states 
with
non-zero condensate number, and they describe the annihilation or creation
of excitations (quasi-particles), or, equivalently, transitions from
an excited state $j$ into the condensate and vice versa. This implies
that the operators $\hat{\alpha}_j,\hat{\alpha}_j^{\dag}$ {\it do not} 
commute
with the condensate operator $\hat{a}_0$. The wave functions $u_j({\bf r})$
and $v_j({\bf r})$ are the corresponding amplitudes associated with
the annihilation of a real particle at position ${\bf r}$, an action
which involves both annihilation or creation of excitations on top of
the condensate.  The time-dependence of the functions $\psi_0,u_j,v_j$
is induced only by the change in the global variables $V_t({\bf
r},t),N_t(t),E_{trap}(t)$ and they are assumed to be time-independent
under the steady-state conditions.

The condensate wave function $\psi_0({\bf r})$ in Eq.~(\ref{psit}) is 
defined
as the solution of the generalised steady-state Gross-Pitaevskii
equation
\begin{equation}
\{{\cal L}_t-\mu+U_0[\bar{N}_0|\psi_0({\bf  r})|^2+2\bar{n}({\bf
r})]\}\psi_0({\bf r}) = 0
\label{GPE}
\end{equation}
where ${\cal L}_t$ is given in Eq.~(\ref{calLt}), while the adiabatic
mean number $\bar{N}_0$ of the condensate atoms and the density
$\bar{n}({\bf r})$ of the non-condensate atoms are calculated
self-consistently by requiring
\begin{equation}
\bar{N}_0+\int d^3 {\bf  r}\bar{n}({\bf  r}) = N_t.
\end{equation}
The functions $u_j({\bf  r}),v_j({\bf  r})$ satisfy the steady-state 
equations
\begin{eqnarray}
\left ( \begin{array}{cc}  {\cal L}_t-\mu+2U_0[\bar{N}_0|\psi_0({\bf  
r})|^2+\bar{n}({\bf  r})] &
U_0 \bar{N}_0 (\psi_0({\bf  r}))^2 \\
-U_0 \bar{N}_0 (\psi_0^*({\bf  r}))^2 & -{\cal L}_t+\mu-
2U_0[\bar{N}_0|\psi_0({\bf  r})|^2+\bar{n}({\bf  r})] \end{array} \right )
\left ( \begin{array}{c} u_j({\bf  r}) \\ v_j({\bf  r}) \end{array} \right )
=E_j\left ( \begin{array}{c} u_j({\bf  r}) \\ v_j({\bf  r}) \end{array} 
\right ) \nonumber \\
+U_0 \bar{N}_0\int d^3 {\bf  r} |\psi_0({\bf  r})|^2
[\psi^*_0({\bf  r})u_j({\bf  r})+\psi_0({\bf  r})v_j({\bf  r})]\left (
\begin{array}{c} \psi_0({\bf  r}) \\
\psi^*_0({\bf  r})\end{array} \right ),
\label{BdG}
\end{eqnarray}
where $E_{j} = \hbar \omega_{j}$ are the $j$th quasi-particle excitation 
energy with respect to the condensate ground state energy, and the second 
term on the right hand side ensures the orthogonality of
the non-condensate functions with the condensate\cite{MorChoiBur97}.
We note that the vectors $\left (
\begin{array}{c} v^*_j \\
u^*_j\end{array} \right )$ satisfy an equation similar to
Eq.~(\ref{BdG}) with $E_j \rightarrow -E_j$; Eqs.~(\ref{GPE})
and~(\ref{BdG}) must be solved self-consistently for any given global
conditions. The mean number of atoms in an excited state
in equilibrium is assumed to be given by the Bose-Einstein distribution
\begin{equation} 
n^{eq}_j=\frac{1}{\exp[\hbar\omega_j/T]-1} \label{BED}.
\end{equation}

In this section we assume a weak output process so that the total
number of atoms in the trap does not change significantly during the
application of the output coupling. Under these conditions the
time-dependence of the operators $\hat{a}_0$ and $\hat{\alpha}_j$ is assumed 
to be simply
\begin{eqnarray}
\hat{a}_0(t)& \approx & \hat{a}_0(0) \label{a_0t} \\
\hat{\alpha}_j(t) & \approx & \hat{\alpha}_j(0)e^{-i\omega_j t}. 
\label{a_jt}
\end{eqnarray}

In our numerical demonstration throughout this paper we take a one 
dimensional
Bose gas of $N_t=2000$ atoms in a harmonic trap with
frequency $\omega$.  The critical temperature for condensation in this
case is $T_c \sim 300\hbar\omega/k$.  We have used a self-consistent
HFB-Popov method to find the wave functions and energies of the
condensate and the excitations. Throughout this paper we take
$U_0=10\hbar\omega\sqrt{2\hbar/m\omega}$. We present calculations for
two temperatures: for $T=10 \hbar\omega/k$ we obtain the chemical
potential $\mu\approx 2.5
\hbar\omega$ and
the non-condensate fraction $\sim 2\%$. At $T=150\hbar\omega/k$ we obtain
$\mu\approx 2.3\hbar\omega/k$ and the non-condensate fraction is $\sim 
44\%$.
We take the interaction strength between the trapped and untrapped atoms
to be $U_1=U_0$. Length will be presented in units of 
$2\sqrt{\hbar/m\omega}$
(``harmonic-oscillator units'').

\subsection{Basic properties of the output}

In order to obtain the properties of the output we expand the output field
operator $\hat{\psi}_f$ in terms of the free modes $\varphi_{{\bf k}}$ in the quasi-steady-state[Eq.~(\ref{psif-expan})].  We
assume $\lambda({\bf r},t)=\lambda({\bf r})e^{-i\Delta_{em}t}$. Using
the form~(\ref{psit}) and~(\ref{dpsit}) of $\hat{\psi}_t$ and the
assumptions~(\ref{a_0t}) and~(\ref{a_jt}), we obtain from
Eq.~(\ref{solbk}) the following equation for the annihilation
operators of the free output modes
\begin{eqnarray}
\hat{b}_{{\bf  k}}(t) & \approx & e^{-i\omega_{{\bf  k}} t} \left \{ 
\hat{b}_{{\bf  k}}(0)
-i \left \{\lambda_{{\bf  k} 0}D_{{\bf  k} 0}(t) \hat{a}_0(0)  
\right.\right.
\nonumber \\
&&\left. \left.
+ \hat{a}_0\frac{1}{\sqrt{\hat{N}_0}}
\sum_j[\lambda_{{\bf  k} j+}D_{{\bf  k} j+}(t) \hat{\alpha}_j(0)
+\lambda_{{\bf  k} j-}D_{{\bf  k} j-}(t) \hat{\alpha}^{\dag}_j(0)]\right\} 
\right\},
\label{solbk1}
\end{eqnarray}
where
\begin{eqnarray} \lambda_{{\bf  k} 0}&=& \int d^3 {\bf  r} \varphi_{{\bf
k}}^*({\bf  r})\lambda({\bf  r})\psi_0({\bf  r})
\label{lmk0} \\
\lambda_{{\bf  k} j+} &=& \int d^3 {\bf  r} \varphi_{{\bf  k}}^*({\bf
r})\lambda({\bf  r})u_j({\bf  r})
\label{lmkjp} \\
\lambda_{{\bf  k} j-} &=& \int d^3 {\bf  r} \varphi_{{\bf  k}}^*({\bf
r})\lambda({\bf  r})v^*_j({\bf  r})
\label{lmkjm}
\end{eqnarray}
are the matrix elements of $\lambda({\bf r})$ between the wave
functions of the collective excitations and the output
states. The time- and energy- dependence is determined by the functions
\begin{equation} 
D_{{\bf
k}\eta}(t)=i\frac{e^{-i(\omega_{out}^{\eta}-\omega_{{\bf  k}})t}-1}
{\omega_{out}^{\eta}-\omega_{{\bf  k}}},
\label{D-def}
\end{equation}
where
\begin{equation}
\hbar\omega_{out}^{\eta}=\mu+\Delta_{em}+E_{\eta},
\end{equation}
with $\eta=0,j+,j-$, and $E_{j+}=E_j$ and $E_{j-}=-E_j$.

With the above definitions, the field operator $\hat{\psi}_f({\bf r},t)$ of
the free atoms can be written as
\begin{eqnarray}
\hat{\psi}_f({\bf  r},t) & \approx & \hat{\psi}_f^{(0)}({\bf  r},t)
-i \Psi_f^0({\bf  r},t) \hat{a}_0(0)  \nonumber \\
&&
-\hat{a}_0(0)\frac{i}{\sqrt{\hat{N}_0}}
\sum_j \left [\Psi_f^{j+}({\bf  r},t) \hat{\alpha}_j(0)
+\Psi_f^{j-}({\bf  r},t) \hat{\alpha}^{\dag}_j(0) \right ],
\label{solpsif}
\end{eqnarray}
where
\begin{equation}
\Psi_f^{\eta}({\bf  r},t)=\sum_{{\bf  k}}\varphi_{{\bf
k}}({\bf  r})
D_{{\bf  k}\eta}(t)\lambda_{{\bf  k}\eta}e^{-i\omega_{{\bf  k}} t}
\label{Psi_f}
\end{equation}
for each $\eta=0,j+,j-$.

This result enables us to calculate various properties of the output
from the trap, assuming Bose-Einstein statistics for the initial
quasiparticle populations inside the trap. For the initial state we
assume that all the cross correlation functions between different
operators $\hat{a}_0, \hat{\alpha}_j, \hat{\alpha}_j^{\dag}$ vanish, and the only non-zero contributions are the
populations
\begin{eqnarray} 
\langle  \hat{a}_0^{\dag} \hat{a}_0 \rangle =N_0\equiv 
n_0^t \label{nt0} \\
\langle \hat{\alpha}^{\dag}_j \hat{\alpha}_j \rangle =n_j\equiv n^t_{j+} 
\label{ntjp} \\
\langle  \hat{\alpha}_j \hat{\alpha}^{\dag}_j \rangle=n_j+1\equiv n^t_{j-}. 
\label{ntjm}
\end{eqnarray}

Any measurable quantities related to the output atoms may be expressed
in terms of the correlation functions of the field operator 
$\hat{\psi}_f({\bf
r},t)$ at different times and space points. In particular, the density
of output atoms at a given point ${\bf r}$ and time $t$ is given by the
equal-time, equal-position correlation function
\begin{equation}
n_{out}({\bf  r},t)=\langle \hat{\psi}_f^{\dag}({\bf  r},t) 
\hat{\psi}_f({\bf  r},t)
\rangle .
\end{equation}
By using Eq.~(\ref{solpsif}) and the
assumptions~(\ref{nt0})-(\ref{ntjm}) we observe that $n_{out}({\bf  r},t)$ 
can be written as a sum over discrete contributions from the levels in the 
trap
\begin{equation}
n_{out}({\bf  r},t)=N_0|\Psi^0_f({\bf  r},t)|^2+\sum_j \left [
n_j |\Psi^{j+}_f({\bf  r},t)|^2+(n_j+1)|\Psi^{j-}_f({\bf  r},t)|^2 \right ],
\end{equation}
where the first term is the condensate output, the second term is the
contribution from the stimulated quantum evaporation of the
thermal excitations, and the third term is the contribution from the
pair-breaking process, as discussed in Ref.~\cite{JapChoiBur99}.

Following similar steps, the number of output atoms  in mode ${\bf 
k}$ of the free
atomic level, $n_{{\bf k}}\equiv\langle \hat{b}_{{\bf
k}}^{\dag} \hat{b}_{{\bf k}} \rangle$, is given by a sum of discrete 
contributions from the different
levels of the trapped gas
\begin{equation}
n_{{\bf  k}}(t)=n_{{\bf  k}}^0(t)+\sum_j \left [
n_{{\bf  k}}^{j+}(t)+n_{{\bf  k}}^{j-}(t) \right ], \label{spect}
\end{equation}
where each term $\eta=0,j+,j-$ in Eq.~(\ref{spect}) has the form
\begin{equation} n_{{\bf  k}}^{\eta}(t)=|\lambda_{{\bf  k}\eta}|^2 |D_{{\bf
k}\eta}(t)|^2 n^t_{\eta}.
\label{spect1}
\end{equation}

The time- and energy- dependence of each of the $n_{{\bf k}}^{\eta}$ is given by the
function
\begin{equation}
|D_{{\bf  k}\eta}(t)|^2=\frac{1}{2}
\left[\frac{\sin[(\omega_{{\bf  k}}-\omega_{out}^{\eta})t/2]}
{(\omega_{{\bf  k}}-\omega_{out}^{\eta})/2} \right]^2.
\label{D2}
\end{equation}
This function has a spectral width in $\omega_{\bf k}$ which decreases with 
time as
$\sim \pi/t$. This spectral width represents the
energy uncertainty dictated by the finite duration of the output
coupling process. The time evolution of the output atoms is therefore
governed by the spectral dependence of the matrix elements
$\lambda_{{\bf k}\eta}$.

In order to analyse the evolution of the output rate and the output
atom density, we define two frequency scales with regard to the matrix elements 
$\lambda_{{\bf  k}\eta}$ for each $\eta=0,j+,j-$: (1) $\Delta\omega_{\eta}$, 
 the
frequency bandwidth within which the matrix
elements $\lambda_{{\bf k}\eta}$, defined in
Eqs.~(\ref{lmk0})-(\ref{lmkjm}), are significant and (2)
$\delta\omega_{\eta}$,  the ``width'' of $\lambda_{{\bf
k}\eta}$ in the $\omega_{{\bf k}} \equiv \hbar k^{2}/2m$ space in the 
vicinity of
$\omega_{{\bf k}}=\omega_{out}^{\eta}$. The weak coupling assumption
is justified if the strength of the coupling, which may be represented
by the parameter $\Lambda=\sqrt{\int d^3 {\bf r}|\lambda({\bf r})|^2}$
is much smaller than $\Delta\omega_{\eta}$ of each trap
state, namely,
\begin{equation}
\Lambda \ll \Delta\omega_{\eta}.
\end{equation}
If this condition is not satisfied, then we expect Rabi oscillations
between the trapped atomic level $| t \rangle$ and the output level $|
f \rangle$\cite{Cohen-Tannoudji}. In the case of weak coupling, we may
identify three temporal regimes:
\begin{enumerate}
\item {\bf Very short times}, $t\ll \Delta\omega_{\eta}^{-1}$. Then the
function $D_{{\bf k}\eta}(t)$ in Eq.~(\ref{D-def}) becomes independent
of ${\bf k}$, and if $\omega_{out}^{\eta}$ lies within the bandwidth
$\Delta\omega_{\eta}$, $D_{{\bf k}\eta}(t)\approx
t$. In this case, the completeness of the set of functions
$\varphi_{{\bf k}}({\bf r})$ implies
\begin{equation} \Psi_f^{\eta}({\bf  r},t)\sim_{t\rightarrow 0} \lambda({\bf
r})\psi_t^{\eta}({\bf  r})t,
\label{Psi-short}
\end{equation}
where $\psi_t^0=\psi_0,\psi_t^{j+}=u_j$, and $\psi_t^{j-}=v_j^*$. The 
initial
shape of the output wave functions before it had time to propagate is
therefore the overlap of the electromagnetic field amplitude and the
corresponding trapped wave function. The density $n_{out}({\bf r})$ in
this case is then similar in shape to the density of atoms in the trap
and the total number of output atoms increases quadratically in
time. This result may also be used to calculate the output beam
immediately after the application of a strong coupling
pulse, before the output beam starts to propagate or Rabi
oscillations occur.

\item {\bf Intermediate times}, $\Delta\omega_{\eta}^{-1}<t<
\delta\omega_{\eta}^{-1}$. In this case the rate of output from each trap
state $\eta$ may show oscillations, which follow from interference
between output from different momentum states.

\item {\bf Long times}, $t\gg \delta\omega_{\eta}^{-1}$. The output from the
internal state $\eta$ is then mainly generated in a narrow range of
energies around $\hbar\omega^{\eta}_{out}$ and the rate of output
$dn_{{\bf k}}/dt$ into these specific modes settles on a constant
value, which is determined by the absolute value of the matrix element
$\lambda_{{\bf k}\eta}$ at $\omega_{{\bf k}}=\omega_{out}^{\eta}$.  It
is then given by the Fermi golden rule
\begin{equation}
\frac{dn_{{\bf  k}}}{dt}=2\pi\sum_{\eta} n^t_{\eta}|\lambda_{{\bf  
k}\eta}|^2
\delta(\omega_{{\bf  k}}-\omega_{out}^{\eta}), \label{Golden}
\end{equation}
and the output rate obtained when one scans the frequency detuning 
$\Delta_{em}$ of the coupling fields measures the magnitude of the matrix 
elements
$|\lambda_{{\bf k}\eta}|^2$ as a function of $\omega_{{\bf k}}$.

\end{enumerate}

The asymptotic behaviour of $|D_{{\bf k}\eta}(t)|^2$ in this limit is
\begin{equation}
|D_{{\bf  k}\eta}(t)|^2 \sim 2\pi\delta(\omega_{{\bf
k}}-\omega_{out}^{\eta})t+
2\frac{P}{(\omega_{{\bf  k}}-\omega_{out}^{\eta})^2},
\label{D2_asymp}
\end{equation}
where the principal part in the second
term is defined as $P/(\omega-\omega^{\eta})^2\equiv \partial/\partial
\omega^{\eta}[P/(\omega-\omega^{\eta})]$. The first term represents a 
linearly increasing mono-energetic
contribution of the level $\eta$ to the output with a constant rate
given in Eq.~(\ref{Golden}). The second term involving the principal part 
represents a
time-independent
non-resonant part that has two physical implications. First, it represents 
those frequency components
that are different from the central resonance frequency which arise due to 
the sudden switching on of the output coupling
field at $t=0$.  Second, this term results in the fields
$\Psi_f^{\eta}(t)$ containing a non-propagating (bound) part
\begin{equation}
(\Psi_f^{\eta}({\bf  r},t))_{bound}=e^{-i\omega_{out}^{\eta} t}P\sum_{{\bf
k}}
\frac{\varphi_{{\bf  k}}({\bf  r})\lambda_{{\bf  k}\eta}}{\omega_{{\bf
k}}-\omega_{out}^{\eta}},
\label{non-prop}
\end{equation}
which stays mainly near the trap. This term appears as a part of the
dressed ground-state of the coupled system, which is a mixture of the
trapped and the untrapped atomic levels. It therefore represents a
virtual transition to the output level while the atoms remain bound to the 
trap.
It is detectable if the atomic detecting system is sensitive enough to
identify small number of atoms in a different Zeeman level near the
main atomic cloud, which would contain atoms in the Zeeman level $| t
\rangle$.  Although this last contribution is in general small
compared to the resonant contributions, it may be significant when
considering the  condensate output ($\eta=0$), which is multiplied
by the large number $N_0$, and therefore may be dominant near the trap
relative to the contributions of the stimulated quantum evaporation (for
$\Delta_{em}<0$) or the pair-breaking (for $\Delta_{em}>0$).  The
condensate contribution can be estimated by $(n^0_f)_{bound}
\approx 2N_0 (\lambda(0)/\omega_{out}^0)^2$, where $\lambda(0)$ is the Rabi
frequency associated with the coupling field at the centre of the
trap.

The spectral widths $\Delta\omega_{\eta}$ and $\delta\omega_{\eta}$
defined above may drastically vary with the structure of the Hilbert
space of the output modes, which is determined by the potential
$V_f({\bf r})$, and also with the spatial shape of the coupling
$\lambda({\bf r},t)$. It is worth mentioning the following limiting
cases:
\begin{enumerate}
\item In the absence of gravity the wave functions $\varphi_{{\bf  k}}$ are
roughly given by the plane waves $e^{i{\bf k}\cdot{\bf r}}$. Then the
matrix elements $\lambda_{{\bf k} 0}$ that couple the condensate to
the free modes is roughly the Fourier transform of the condensate
wave function $\psi_0({\bf r})$. If no momentum kick is provided, then
their width in momentum space is given in terms of the spatial width
$r_0$ of the condensate by $\delta k_0\sim 1/r_0$. The corresponding spectral width,
$\Delta\omega_0$, is then 
\begin{equation} 
\Delta\omega_0\sim \delta\omega_0 \sim  \hbar /2mr_0^2 <
\omega, 
\end{equation}
which implies that the time it will take to achieve a constant rate of
output from the condensate is larger than the period of the trap. If the
condensate is broadened by a strong collisional repulsion then this time may 
be
much greater than this period, which is typically of the order of $10$ ms.
\item In the case where a momentum kick ${\bf k}_{em}$ is provided,  the spectral width for the condensate output becomes
\begin{equation}
\delta\omega_0\sim \hbar |{\bf k}_{em}| \delta k_0/2m \sim \hbar |{\bf k}_{em}| /2mr_0.
\end{equation}
This makes the time for achieving steady output shorter by a factor
$(|{\bf k}_{em}| r_0)^{-1}$ compared to the previous case. If $|{\bf k}_{em}|$
corresponds to an optical wavelength then this factor may be of the
order of 10.
\item In the presence of gravity the spectral width of $\lambda_{{\bf
k}\eta}$ is determined mainly by the gradient of the gravitational
potential over the spatial extent of the corresponding wave function
$\psi_t^{\eta}$.  A typical value of this gradient for the condensate
wave function in the experiment of Ref.~\cite{Munich99} is about
$\delta\omega_0\sim 2\pi \times 10$ kHz.  In this case the time needed
to achieve a steady output is much shorter, in the order of
$\sim 0.1$ ms.
\end{enumerate}

\begin{figure}[t] \begin{center}
\psfig{width=12cm,file=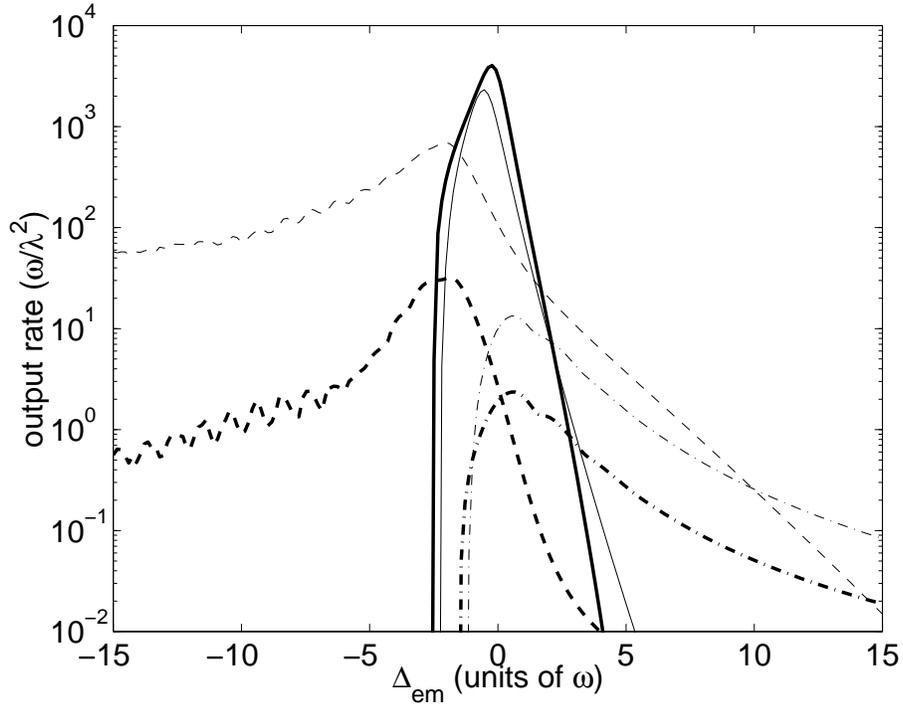}
\end{center}
\caption{\protect \footnotesize
The rate of output as a function of $\Delta_{em}$ for atoms emerging from 
various different processes at temperatures $T=10 \hbar\omega/k$
(bold line) and $T=150 \hbar\omega/k$ (thin line).
Solid line: $dn^0_f/dt$, output from the condensate component; Dashed line: 
$\sum_j dn^{j+}_f/dt$, from stimulated quantum
evaporation; Dash-dotted line: $\sum_j dn^{j-}_f/dt$, from pair-breaking.
A constant density of states $\rho(\omega_{{\bf  k}})=1$ was used.}

\label{fig:spect}
\end{figure}

The rate of transfer of atoms into the output level as a function of
$\Delta_{em}$ in our one-dimensional example is plotted in
Figure~\ref{fig:spect}. This rate is a sum of contributions from the
condensate and the excited states in the trap
\begin{equation}
\frac{dN_{out}}{dt}=\frac{dn_f^0}{dt}+\sum_j\left[\frac{dn^{j+}_f}{dt}
+\frac{dn^{j-}_f}{dt}\right], \end{equation}
where
\begin{equation} 
n^{\eta}_f=\int d^3 {\bf  r} n^{\eta}_{out}({\bf
r})=\sum_{{\bf  k}} n^{\eta}_{{\bf  k}}.
\label{N_out} \end{equation}
The rate of output from the condensate component, $dn^0_f/dt$ (solid line),
that from stimulated quantum evaporation, $\sum_j dn^{j+}_f/dt$ 
(dashed
line), and finally from pair-breaking, $\sum_j dn^{j-}_f/dt$ (dash-dotted
line) are shown for temperatures $T=10 \hbar\omega/k$ ($\sim 0.03
T_c$, bold line) and $T=150 \hbar\omega/k$ ($\sim 0.5 T_c$, thin
line), in the case where no momentum is transferred from the EM field
(${\bf k}_{em}=0$). The threshold below which the condensate output is not produced is
at $\Delta_{em}=-\mu$, which is slightly different for the two
temperatures.  To prevent unphysical effects that follow from the
divergence of the density of states at small momenta in
one-dimensional systems, we have assumed that the density of momentum
states per energy is constant, $\rho(\omega_{{\bf k}})=1$.  As to be anticipated, 
the
composition of the output beam changes as a function of $\Delta_{em}$:
for negative values of $\Delta_{em}$ the dominant contribution is from the 
stimulated
quantum evaporation of initially excited levels in the trap; for
positive $\Delta_{em}$ it is the contribution of pair-breaking  that dominates the output.  The output contains overwhelmingly condensate components at central values of $\Delta_{em}$.  Comparison of the
results for $T=10\hbar\omega/k$ and $T=150\hbar\omega/k$ shows that output
rate from the pair-breaking is pronounced mainly at low temperatures.

Figure~\ref{fig:density} is a one-dimensional demonstration of the
output density for coupling frequencies in (a) the stimulated quantum
evaporation regime ($\Delta_{em}=-5\omega$), (b) the coherent
output regime ($\Delta_{em}=0$), and (c) the pair-breaking regime
($\Delta_{em}=8\omega$).  At very short times the output density from
each level has the same shape as the density of the given level in the
trap, as follows from Eq.~(\ref{Psi-short}). After a short time, the
output atoms emerge mainly in two momentum states $\varphi_{{\bf k}}$
corresponding to the right- and left-propagating waves
with energy $\hbar\omega_{out}^{\eta}$.
Since the magnitude of the matrix
elements $\lambda_{{\bf k}\eta}$ for these two modes are equal, the
output beam corresponding to a given component $\eta$ forms a standing
wave and consequently the density $n_{out}^{\eta}(x)$ becomes
oscillatory. This aspect is demonstrated below, when we discuss the
coherence of the output. These standing waves are not expected in cases 
where
the inversion symmetry is broken, such as the case with ${\bf k}_{em} \neq 
0$. In cases (a) and (c), where $\Delta_{em}$ is
very positive or negative, one can see that the output density from the 
condensate has a steady component that remains near the trap. This part 
corresponds to the appearance of the bound states discussed above in conjunction with Eq.~(\ref{non-prop}).

\subsection{Coherence of the output}

The concept of the $n$-th order coherence in a quantum system was
originally developed in the optical context to quantify the correlations
in the field\cite{Glauber}.  The first order
coherence measures the fringe contrast in a typical Young's double slit
experiment, while the second order coherence gives indications of
counting statistics in, say, Hanbury-Brown-Twiss experiments.  A theory of
the coherence of matter-waves was presented only
recently\cite{Goldstein98} and the case of a trapped
Bose gas has also been discussed\cite{Glauber99}. It follows that, for matter waves,
the theory of coherence which is applicable to real
experiments is much more complicated than that for the optical case.
However, any measures of coherence must involve correlation functions
between the matter-wave field operators. For simplicity, we use here
definitions of matter wave coherence functions that are equivalent to
the optical definitions by replacing the usual electromagnetic field
operators by the matter-wave field operators.

\subsubsection{First-order coherence}

The first order coherence function $g_f^{(1)}({\bf r},{\bf r'},t,t')$ for 
the
output atoms is defined as
\begin{equation}
g_f^{(1)}({\bf r},{\bf r'},t,t')  =  \frac{\langle \hat{\psi}^{\dag}_f({\bf
r},t)\hat{\psi}_f({\bf r'},t')\rangle}{\sqrt{\langle 
\hat{\psi}^{\dag}_f({\bf
r},t)\hat{\psi}_f({\bf r},t)\rangle \langle \hat{\psi}^{\dag}_f({\bf
r'},t')\hat{\psi}_f({\bf r'},t')\rangle}},
\end{equation}
where $g^{(1)} = 1$ implies full coherence and $g^{(1)} = 0$ implies
total incoherence. The first-order coherence for a random or thermal
mixture of many modes typically takes the maximal value for ${\bf
r}={\bf r}'$ (i.e.  $g^{(1)}({\bf r},{\bf r}) = 1$) and falls down to
zero for large $|{\bf r}-{\bf r}'|$ or $|t-t'|$.  Highly monochromatic
beams, however, are characterised by the fact that $g^{(1)}=1$ even
for large $|{\bf r}-{\bf r}'|$ or $|t-t'|$, implying high fringe
visibility even if widely separated parts of the beam interfere.

\begin{figure}[t] \begin{center}
\psfig{width=12cm,file=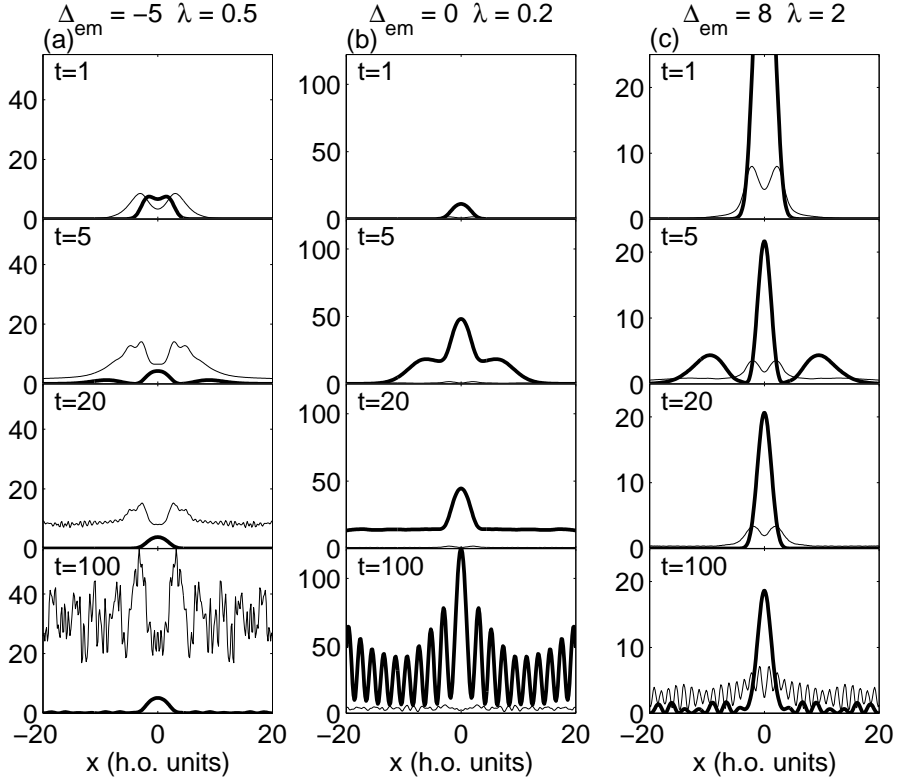} 
\end{center}
\caption{\protect \footnotesize
A one-dimensional demonstration of the temporal evolution of the coherent 
component (bold line)
and the thermal component (thin line) of the output atomic density at $T=150 
\hbar\omega/k$ ($\sim 0.5T_c$) for different coupling strengths and 
detunings.
(a) $\Delta_{em}=-5\omega$, $\lambda = 0.5\omega$, (b) $\Delta_{em}=0$,  
$\lambda = 0.2\omega$, and (c)
$\Delta_{em}=8\omega$,  $\lambda = 2\omega$.
The output density from the condensate has a steady component that remains
near the trap; this part corresponds to the appearance of the bound states
discussed after Eq.~(\ref{non-prop}).}
\label{fig:density}
\end{figure}

\begin{figure}[t] \begin{center}
\psfig{width=12cm,file=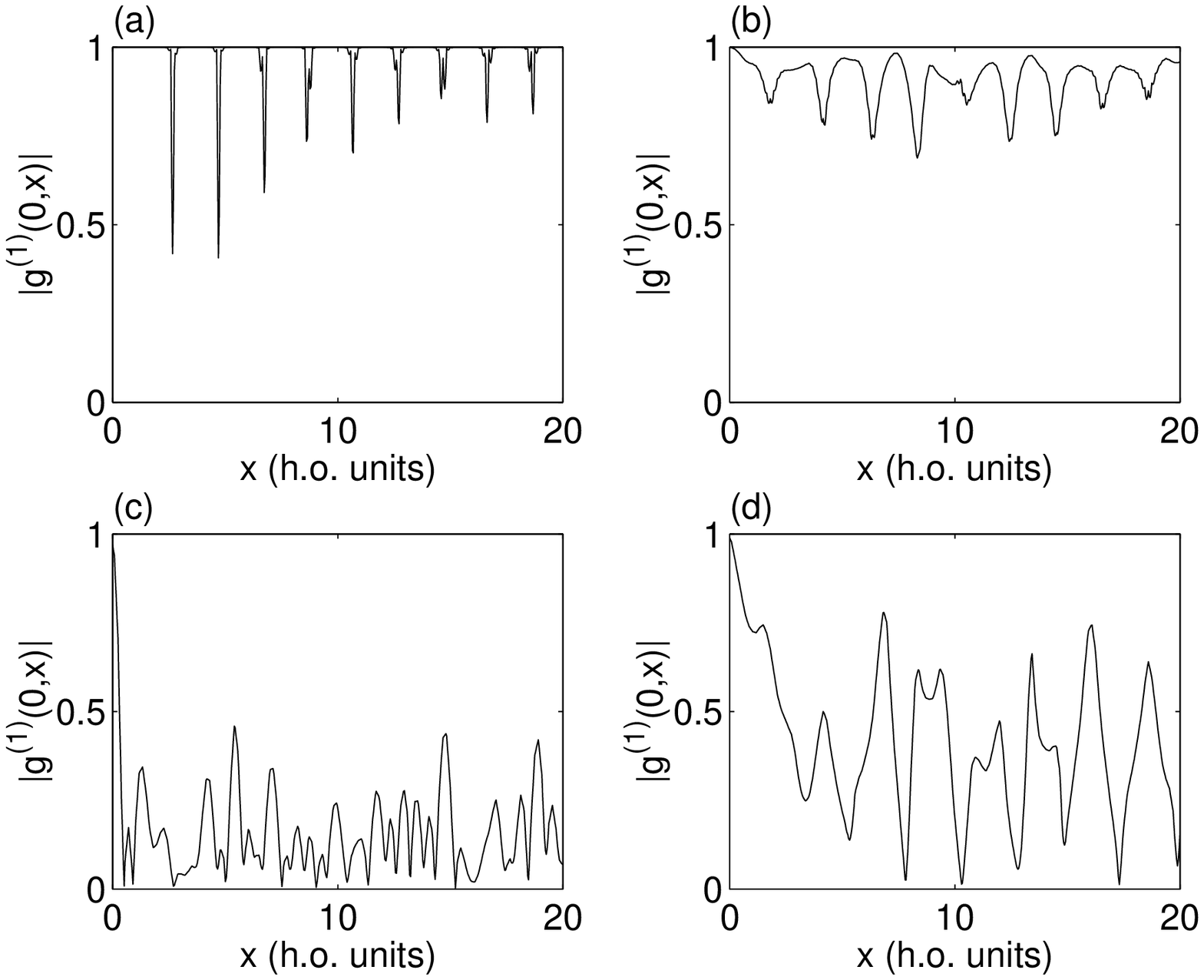}
\end{center}
\caption{\protect \footnotesize
The first order coherence $g_f^{(1)}(x_1,x_2,t)$ of the output atoms
as a function of $x=x_2$ for a fixed value of $x_1=0$
at time $t=100/\omega$, for (a) $T=10 \hbar\omega/k, \Delta_{em}=0$ 
(Dominant
coherent output), (b) $T=150\hbar\omega/k$, $\Delta_{em}=0$, (c)
$T=150\hbar\omega/k, \Delta_{em}=-5\omega$ (dominant thermal output) and (d)
$T=150\hbar\omega/k, \Delta_{em}=8\omega$ (dominant pair-breaking).}
\label{fig:g1}
\end{figure}

An atomic beam weakly coupled out from a finite temperature Bose-gas is, in general,
a mixture of quasi-monochromatic beams originating from the condensate
and the internal excitations in the trap.  The nature of this mixture 
depends on the frequency, shape and
momentum transfer from the electromagnetic field, and correspondingly
the coherence properties are significantly affected.  Following the
quasi-steady-state solution for $\hat{\psi}_f({\bf r})$ in
Eq.~(\ref{solpsif}) we find for our first order coherence,
\begin{equation}
g_f^{(1)}({\bf r},{\bf r'},t,t')   =
\frac{1}{\sqrt{n_{out}({\bf r})n_{out}({\bf r'})}}
\sum_{\eta=0,j+,j-}
N_{\eta}[\Psi_f^{\eta}({\bf r},t)]^* \Psi_f^{\eta}({\bf r'},t').
\end{equation}
The coherence is maximal if only one of the terms from the sum over
$\eta$ is dominant. Fig.~\ref{fig:g1} shows the first order coherence
function $g_f^{(1)}(x_1,x_2,t)$ of the output atoms in our
one-dimensional demonstration as a function of $x_2$ for fixed 
$x_1=0$ and time $t=100/\omega$. When $\Delta_{em}=0$ and the
temperature is low ($T=10 \hbar\omega/k$, Fig.~\ref{fig:g1}a) the
coherence function is unity except for points where the condensate
density vanishes (see Fig.~\ref{fig:density}b). At $T=150
\hbar\omega/k$ (Fig.~\ref{fig:g1}b) the thermal component is larger
and it is more dominant near the points where the density of the
condensate component is low. These features are unique to configurations
where the output has a form of standing
matter-waves. When $\Delta_{em}=-5\omega$ (Fig.~\ref{fig:g1}c) the
thermal components are dominant (see Fig.~\ref{fig:density}a) and
the coherence drops much lower than unity.  When $\Delta_{em}=8\omega$
(fig.~\ref{fig:g1}d) only few thermal output components exist from the 
pair-breaking process,  and consequently one obtains a comparatively high coherence function.

\subsubsection{Second-order coherence}

Of particular interest are the intensity correlations of the fields  which 
are
important, for example, in experiments involving non-resonant light
scattering from an atomic gas\cite{Javanainen99}.  These intensity
correlations are expressed in terms of the second order correlation
function, $g_f^{(2)}({\bf r}_1,{\bf r}_2,t_1,t_2)$, which is defined as
\begin{equation}
g_f^{(2)}({\bf r}_{1}, t_{1}; {\bf  r}_{2}, t_{2})
=  \frac{\langle \hat{\psi}^{\dag}_f({\bf r}_{1}, t_{1})
\hat{\psi}^{\dag}_f({\bf r}_{2}, t_{2})  \hat{\psi}_f({\bf r}_{1}, t_{1})
\hat{\psi}_f({\bf r}_{2},
t_{2})\rangle}{\langle \hat{\psi}^{\dag}_f({\bf r}_{1}, t_{1})
\hat{\psi}_f({\bf
r}_{1}, t_{1}) \rangle  \langle \hat{\psi}^{\dag}_f({\bf r}_{2}, t_{2})
\hat{\psi}_f({\bf r}_{2}, t_{2}) \rangle}.
\end{equation}
The function measures the joint probability of detecting two
atoms at two space-time points. If the detection probability of an
atom is independent of the detection probability of another atom then
$g^{(2)}=1$ and the probability distribution is Poissonian. This is the
case for a coherent state of the matter field. However, for a thermal
state the correlation function at the same space-time point is 
$g_f^{(2)}({\bf
r}_{1}={\bf r}_{2},t_{1}=t_{2})\approx 2$.  This implies that the atoms are 
``bunched,'' i.e. there is a larger probability to detect two atoms 
together.

The
second-order correlation function at equal position and time points ${\bf r}_{1} = {\bf r}_{2}$ and $t_{1} = t_{2}$ was previously
calculated for a trapped Bose gas\cite{DodBurEdw97}. Here we shall
follow the same treatment for calculating the second-order coherence
of the output beam. We decompose the field operator $\hat{\psi}_f({\bf r}, 
t)$ into a
part proportional to the condensate and a part proportional to the
excited states in the trap, and apply Wick's theorem to the expectation
value of the product of four non-condensate operators:
\begin{equation}
\langle \hat{\psi}_{nc}^{\dag}({\bf  r})\hat{\psi}_{nc}^{\dag}({\bf  
r})\hat{\psi}_{nc}({\bf
r})\hat{\psi}_{nc}({\bf r})
\rangle
= 2 \tilde{n}^{2}({\bf r}) + \tilde{m}^*({\bf r}) \tilde{m}({\bf r}),
\end{equation}
where we have defined $\tilde{n}({\bf r}) = \langle \hat{\psi}_{nc}^{\dag} ({\bf   r}) \hat{\psi}_{nc}({\bf r}) \rangle$ and
$\tilde{m}({\bf r}) = \langle \hat{\psi}_{nc}({\bf  r}) \hat{\psi}_{nc}({\bf 
  r}) \rangle$.
One then obtains for the second order
coherence of the output atoms:
\begin{equation}
g_f^{(2)}({\bf r},t) = 1 + \frac{1}{n_{out}^{2}({\bf r})} \left \{
2 {\rm Re} \left[n_{out}^0({\bf r})\tilde{n}_{out}({\bf r})
+ [(\Psi_{out}^0({\bf r}))^*]^{2}\tilde{m}_{out}({\bf r})
\right] + \tilde{n}_{out}^2({\bf r}) +  |\tilde{m}_{out} ({\bf r})|^{2}
\right\},
\end{equation}
where $\tilde{n}_{out}({\bf r})=\sum_j[n_{out}^{j+}({\bf
r})+n_{out}^{j-}({\bf r})]$ and $\tilde{m}_{out}({\bf r})=\sum_j
\Psi_f^{j+}\Psi_f^{j-}(2n_j+1)$.

We note that although in Ref.~\cite{DodBurEdw97} the terms proportional
to $\tilde{m}$ had negligible contribution, here they may play an
important role even at zero temperature in situations where the tuning of the coupling EM-field frequency yields an output beam that emerges mainly from the
non-condensate parts of the trapped gas.

\begin{figure}[t] \begin{center}
\psfig{width=12cm,file=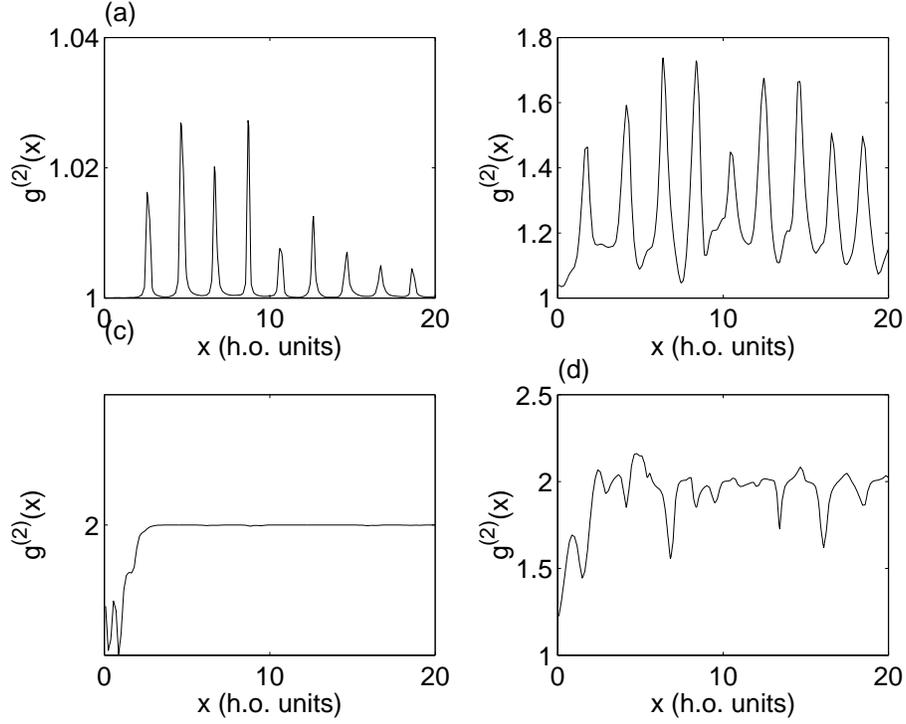} 
\end{center}
\caption{\protect \footnotesize
Equal-time single-point second order coherence function $g_f^{(2)}(x)$ of 
the
output atoms at time $t=100/\omega$. Figures (a)-(d) correspond to the same 
cases
plotted in figure~\protect\ref{fig:g1} of the first-order coherence.}
\label{fig:g2}
\end{figure}

The equal-time, single-point intensity correlation of the output
atoms after a time $t=100/\omega$ in few typical cases is shown in
Fig.~\ref{fig:g2}. If the output condensate is dominant
(Fig.~\ref{fig:g2}a) the function $g^{(2)}(x)$ is equal to unity
except at discrete points where the output condensate wave function
vanishes.  At a higher temperature (Fig.~\ref{fig:g2}b) the thermal
output components tend to raise $g^{(2)}(x)$ near the points where the
coherent part is small. In the case where the thermal component is
dominant ($\Delta_{em}=-5\omega$, Fig.~\ref{fig:g2}c) $g^{(2)}(x)$
assumes the value of 2. In the case where the pair-breaking is dominant
(Fig.~\ref{fig:g2}d) the intensity correlations tend to assume values
greater than 2. This can be interpreted as an atom bunching effect
caused by the combination of the process of pair-breaking with the
stimulated quantum evaporation of the thermal states.

\section{Evolution of the trapped gas}
\label{sec:dynamics}

The last section was devoted to a discussion of the properties of the
output in the quasi-steady-state approximation, where the Bose gas in
the trap is assumed to remain unchanged by the output coupling
process. In this section we describe the internal dynamics of the
trapped atomic Bose gas during the output coupling process. During
this process, the trapped atomic population of the condensate state
and each of the excited states change in a different way and the
system is driven out of equilibrium.  In the typical case where the
duration of the coupling process is short compared to the duration of
relaxation processes at very low temperatures\cite{Damping0,Damping1},
the dynamics is represented by approximate solutions of
Eq.~(\ref{eq:psi0}).  In the case of weak coupling, the solutions are
best represented in terms of the adiabatic basis of the system, which
are the steady-state HFB-Popov solutions for a given total number of
particles and given total energy of the system. It can serve as a good
basis as long as the changes in the conditions in the trap are slow
enough compared with the trap frequency.

We begin by first introducing a two-component vector formalism that is
convenient for dealing with the many modes of excitations.  We then
obtain linear equations of motion for the creation and annihilation
operators of the condensate and excitations in the adiabatic basis. In the 
adiabatic conditions these equations may be simplified and solved
analytically. A perturbation solution is then presented, which is
suitable for describing the short time evolution. Finally, we find that the number-conserving formalism fails to describe the evolution of the condensate
number in the pair-breaking regime. This problem is discussed and
cured in the end of this section.

\subsection{Vector formalism for the trapped atomic gas}

The dynamics of the excited states in the trap is usually described by a
set of two coupled equations of the form Eq.~(\ref{BdG}), which was
discussed in Sec.~\ref{sec:out}, or its time-dependent
version\cite{CasDum98}. This form, as well as the fact that the
expansion for the field operator in Eq.~(\ref{psit}) involves
the quasiparticle creation and annihilation operators 
$\hat{\alpha}_j,\hat{\alpha}_j^{\dag}$,
motivates the introduction of the two-component vector
formalism as follows.

First, we define the normalised condensate operators
\begin{equation}
\hat{c}_0 = \hat{a}_0\frac{1}{\sqrt{\hat{N}_0}}; \
\hat{c}^{\dag}_0=\frac{1}{\sqrt{\hat{N}_0}}\hat{a}_0^{\dag},
\label{c0} \end{equation}
which are well-defined in the space spanned by states with non-zero
condensate number; within this space, they satisfy
$\hat{c}_0 \hat{c}_0^{\dag} = \hat{c}_0^{\dag} \hat{c}_0=1$.  We then define 
the two-component column vector operator
\begin{equation}
\hat{\xi}_t({\bf  r},t)=\left ( \begin{array}{c}
\hat{c}_0^{\dag} \hat{\psi}_t({\bf r},t) \\ \hat{\psi}_t^{\dag}({\bf r},t) 
\hat{c}_0 \end{array}
\right )
\label{def:xi}
\end{equation}
which describes transitions from the condensate state to itself and to
and from the excited states.

The expansion of $\hat{\psi}_t({\bf r},t)$ in Eq.~(\ref{psit}) 
and~(\ref{dpsit}) is
equivalent to the expansion of $\hat{\xi}_t({\bf r}, t)$ in terms of the
two-component wave function vectors as
\begin{equation}
\hat{\xi}_t({\bf  r},t)=\sum_{\eta=-\infty}^{\infty} \xi_{\eta}({\bf
r},t)\hat{\alpha}_{\eta}(t)
e^{-i\int_0^t dt' E_{\eta}(t')}, \label{xi-expan}
\end{equation}
where the index $\eta$ takes integer values from $-\infty$ to
$\infty$. The index $\eta=0$ corresponds to the condensate;
negative indices stand for solutions of Eq.~(\ref{BdG}) with negative
energies $E_{-j}=-E_j$, and operators such that
$\hat{\alpha}_{-j}=\hat{\alpha}_j^{\dag}$. In addition, we note that  the condensate operator
\begin{equation}
\hat{\alpha}_0\equiv \hat{c}_0^{\dag}\hat{a}_0=\sqrt{\hat{N}_0}
\end{equation}
is Hermitian. The two-component vectors 
$\xi_{\eta}({\bf r}, t)$ are defined as
\begin{equation}
\xi_0\equiv\left ( \begin{array}{c} \psi_0 \\
\psi^*_0\end{array} \right ),\; \xi_j\equiv\left ( \begin{array}{c} u_j \\
v_j\end{array} \right ), \;
\xi_{-j}\equiv \left ( \begin{array}{c} v^*_j \\ u^*_j\end{array} \right ).
\end{equation}
 The time dependence of the vectors $\xi_{\eta}({\bf r},t)$ 
and the
energies $E_{\eta}(t)$ in Eq.~(\ref{xi-expan}) is governed by the
time-dependence of the global variables of the system, while the time
dependence of the coefficients $\hat{\alpha}_{\eta}$, $\eta \in (-\infty, 
\infty)$, represents changes
in the populations of the condensate and the excited states.

The usual orthogonality and normalisation conditions for the
eigenfunction $u_j$ and $v_j$ are written in the vectorised notation
as
\begin{equation} \int d^3 {\bf  r} \xi_0^{\dag}({\bf  r})\xi_{\eta}({\bf
r})=\int d^3 {\bf r}
\xi^{\dag}_0({\bf  r})\sigma_3\xi_{\eta}({\bf  r})=0, \label{ortho0}
\end{equation}
\begin{equation} \int d^3 {\bf  r}
\xi_{\eta}^{\dag}({\bf  r})\sigma_3\xi_{\nu}({\bf  r})={\rm sign}
\{E_{\eta}\}\delta_{\eta\nu},
\label{ortho}
\end{equation}
for any $\eta,\nu\neq 0$. Here
\begin{equation}
\xi^{\dag}_j\equiv (u^*_j\ v^*_j);\ ;\ \xi^{\dag}_{-j}\equiv
(v_j\ u_j)
\end{equation}
are the two component row vectors and $\sigma_3$ denotes a $2 \times 2$ matrix,
\begin{equation}
\sigma_3=\left ( \begin{array}{cc} 1 & 0 \\ 0 & -1
\end{array} \right ).
\end{equation}

\subsection{Equations of motion for the operators}

We now derive the equations of motion for the operators
$\hat{\alpha}_{\eta}$ corresponding to transitions from the condensate to
the adiabatic eigenmodes of the system and vice versa. We first
multiply Eq.~(\ref{eq:psi0}) by $\hat{c}^{\dag}_0 e^{i\Phi}$.  The resulting
equation, together with its Hermitian conjugate, form a set of
equations which can be expressed in the following vector form
\begin{eqnarray}
\dot{\hat{\xi}}_t &=& (\dot{\hat{\xi}}_t)^{(0)} - \int_0^t dt' \int d^3
{\bf  r}'
e^{i\sigma_3[\Phi(t)-\Phi(t')]}\tilde{G}({\bf  r},{\bf  r}',t,t')
\hat{\xi}_t({\bf  r}',t') \nonumber \\
&& -i\tilde{\lambda}^{\dag}({\bf  r})\sigma_3 \hat{\xi}_f^{(0)}({\bf  r})
\label{eq:xit}
\end{eqnarray}
where $\tilde{G}({\bf  r},{\bf  r}',t,t')$ and $\tilde{\lambda}({\bf  r})$ are the matrices 
\begin{equation}
\tilde{G}({\bf  r},{\bf  r}',t,t')=\left ( \begin{array}{cc}
\hat{c}_0^{\dag}(t)G({\bf  r},{\bf  r}',t,t')\hat{c}_0(t') & 0 \\
0 & \hat{c}_0^{\dag}(t')G^*({\bf  r},{\bf  r}',t,t')\hat{c}_0(t) \end{array} 
\right ),
\end{equation}
\begin{equation}
\tilde{\lambda}=\left ( \begin{array}{cc} \lambda({\bf  r}) & 0 \\ 0 &
\lambda^{*}({\bf  r}) \end{array} \right ),
\end{equation}
and in a similar manner to $\hat{\xi}_t({\bf r},t)$ of Eq.~(\ref{def:xi}),
\begin{equation}
\hat{\xi}_f^{(0)}({\bf r}, t) = \left (\begin{array}{c} 
\hat{c}_0^{\dag}\hat{\psi}_f^{(0)}({\bf r}, t)\\
(\hat{\psi}_f^{(0)})^{\dag}({\bf r}, t) \hat{c}_0\end{array} \right ).
\end{equation}
$\hat{\xi}_f^{(0)}({\bf r}, t)$ describes the free evolution of the output 
field operator
$\hat{\psi}_f({\bf r}, t)$, as given in
Eqs.~(\ref{psif0}),~(\ref{psif0b}).  The term $(\dot{\hat{\xi}}_t)^{(0)}$ is
the operator describing the free evolution of $\hat{\xi}_t$ inside the trap 
in
the absence of the output coupling but with a given adiabatic change in
the global variables. Here we use the same approximations as in
Eqs.~(\ref{a_0t}),~(\ref{a_jt}), which is equivalent to
\begin{equation}
(\dot{\hat{\xi}}_t)^{(0)}({\bf  r},t)=-i\sum_{\eta} E_{\eta}
\xi_{\eta} \hat{\alpha}_{\eta}e^{-i\int_0^t E_{\eta}(t') dt'}.
\end{equation}
The time derivative of $\dot{\hat{\xi}}_t$ on the left hand side of
Eq.~(\ref{eq:xit}) may then be written as
\begin{equation}
\dot{\hat{\xi}}_t=\sum_{\eta} e^{-i\int_0^t dt' E_{\eta}(t')}
[\dot{\xi}_{\eta} \hat{\alpha}_{\eta} + \xi_{\eta}\dot{\hat{\alpha}}_{\eta}
-iE_{\eta}\xi_{\eta} \hat{\alpha}_{\eta}]. \label{dot-xi}
\end{equation}
The first term corresponds to the time dependence due to the change
in the global variables, the second term is due to the change in the
populations of the condensate and excited states, while the third term
cancels with $(\dot{\hat{\xi}}_t)^{(0)}$ on the right-hand side of
Eq.~(\ref{eq:xit}). We multiply Eq.~(\ref{eq:xit}), in turn, by
$\xi_{\eta}^{\dag}\sigma_3$ for every $\eta\neq 0$ and by
$\frac{1}{2}\xi_0$ for $\eta=0$, and integrate over ${\bf r}$. This
multiplication should be understood as an inner product between row
and column vectors.  By applying the orthogonality and normalisation
relations in Eq.~(\ref{ortho0}) and Eq.~(\ref{ortho}) we obtain the
required equation of motion
\begin{equation}
\dot{\hat{\alpha}}_{\eta}  = -\sum_{\nu} M_{\eta\nu}(t) 
\hat{\alpha}_{\nu}(t)
-\sum_{\nu}\int_0^t dt' G_{\eta\nu}(t,t') \hat{\alpha}_{\nu}(t')
-i\int d^3 {\bf  r} F_{\eta}({\bf  r},t) \hat{\xi}^{(0)}_f({\bf  r},t).
\label{eq:amu}
\end{equation}
Here
\begin{equation}
M_{\eta\nu}(t)=e^{i\int_0^t dt' [E_{\eta}(t')-E_{\nu}(t')]}
\int d^3 {\bf  r} \xi^{\dag}_{\eta}({\bf  r}) \sigma_{\eta}
\dot{\xi}_{\nu}({\bf  r})
\end{equation}
is a matrix with zero diagonal, which describes mixing between the
adiabatic levels that is induced by the change in the global
variables.  This term in Eq.~(\ref{eq:amu}) may be neglected in the
adiabatic limit where the change in the global variables is very
slow. Its effect in slightly non-adiabatic conditions will be
discussed elsewhere\cite{JapBand99}.

The second and third terms in Eq.~(\ref{eq:amu}) describe changes in
the trap which are directly induced by the output coupling.
Defining
\begin{equation}
\Phi_{\eta}(t)=\int_0^t dt' [\mu(t')+\sigma_3 E_{\eta}(t')],
\end{equation}
and
\begin{equation}
\sigma_{\eta}=\left\{\begin{array}{cc} \sigma_3 & \eta>0 \\ \frac{1}{2} &
\eta=0 \\
-\sigma_3 & \eta<0 \end{array} \right. ,
\end{equation}
one may write
\begin{equation}
G_{\eta\nu}(t,t')  = \int d^3 {\bf  r} \int d^3 {\bf  r}'
\xi_{\eta}^{\dag}({\bf  r},t)\sigma_{\eta}
\tilde{G}({\bf  r},{\bf  
r}',t,t')e^{i\sigma_3[\Phi_{\eta}(t)-\Phi_{\nu}(t')]}
\xi_{\nu}({\bf  r}',t'),  \label{Gamma_mat}
\end{equation}
and  
\begin{equation} F_{\eta}({\bf  r})  =  \xi_{\eta}^{\dag}({\bf
r})\sigma_{\eta}\sigma_3
e^{i\sigma_3\Phi_{\eta}}
\tilde{\lambda}^*({\bf  r})
\end{equation}
which describes the effect of the zero-field fluctuations.
An exact analytical solution of Eq.~(\ref{eq:amu}) is, in general, not
possible. However, in the following we present two methods of
approximate solutions to this equation: an adiabatic approximation,
which is suitable for describing the evolution at long enough times,
and a perturbative expansion, which is suitable for short times.

\subsection{Solution via adiabatic approximation}

First we consider the adiabatic and quasi-continuous case where the
functions $\xi_{\eta}$ change very slowly with time and the coupling
amplitude is given by $\lambda({\bf r},t)=\lambda({\bf
r})e^{-i\Delta_{em}t}$. In this case we let $M_{\eta\nu}
\approx 0$,
Second, Eq.~(\ref{eq:amu}) is further simplified by finding an
approximate expression for the integral involving $G_{\eta\nu}(t,t')$.
We make a Markovian approximation, which transforms the integro-differential
equation~(\ref{eq:amu}) into an ordinary differential equation, which can 
then
be solved analytically.
Following the definition of $G({\bf  r},{\bf  r}',t,t')$ in terms of the
free output modes denoted by ${\bf  k}$ [Eq.~(\ref{Gfk}) and~(\ref{Gamma})],
the functions $G_{\eta\nu}(t,t')$ may be written as
\begin{equation}
G_{\eta\nu}(t,t') =  \sum_{{\bf  k}} \bar{\lambda}^{\dag}_{{\bf  k}\eta}(t)
\sigma_{\eta}e^{-i[\omega_{{\bf  k}}-\Delta_{em}]\sigma_3(t-t') }
e^{i[\Phi_{\eta}(t)-\Phi_{\nu}(t')]}
\bar{\lambda}_{{\bf  k}\nu}(t'), \label{Gmunu0}
\end{equation}
where
\begin{equation}
\bar{\lambda}_{{\bf  k}\eta}(t) = \left ( \begin{array}{c} \lambda_{{\bf
k}\eta} \\ \lambda^*_{{\bf k},-\eta}\end{array} \right ),
\end{equation}
with the matrix element $\lambda_{{\bf k}\eta}$ as defined in
Eqs.~(\ref{lmk0})-(\ref{lmkjm}). The time dependence of the matrix
elements $\lambda_{{\bf k}\eta}$ is induced only by the change in the
global variables, which is assumed to be slow. The sum over ${\bf k}$
in Eq.~(\ref{Gmunu0}) may then be regarded as a Fourier transform of
the products $\bar{\lambda}^{\dag}_{{\bf
k}\eta}\sigma_{\eta}\bar{\lambda}_{{\bf k}\nu}$ over $\omega_{{\bf
k}}$ at ``fixed'' point in time $\tau\equiv t-t'$. The width 
$\Delta\tau_{\eta\nu}$ of
$G_{\eta\nu}(\tau)$ as a function of $\tau$ is then roughly given by
the inverse of the spectral width $\Delta\omega_{\eta\nu}$ of the
product of the matrix elements $\lambda_{{\bf k}\eta}$ and
$\lambda_{{\bf k}\nu}$, which is, in turn, given by the smallest of
the spectral widths $\Delta\omega_{\eta}$ and $\Delta\omega_{\nu}$
of the corresponding matrix elements. In the same conditions that
allow the weak coupling approximations done in Eq.~(\ref{Golden}),
i.e., when $G_{\eta\nu}\Delta\tau_{\eta\nu} \ll 1$ and $t\gg
\Delta\tau_{\eta\nu}$, we may take $\hat{\alpha}_{\eta}(t')\approx
\hat{\alpha}_{\eta}(t)$ in Eq.~(\ref{eq:amu}) and 
$\hat{c}_0^{\dag}(t)\hat{c}_0(t')=1$,
and extend the integration over $t'$ to $-\infty$, namely
\begin{equation}
\int_0^t dt' G_{\eta\nu}(t,t') \hat{\alpha}_{\nu}(t')\approx
\Gamma_{\eta\nu}(t)e^{i\int_0^t dt'
[E_{\eta}(t')-E_{\nu}(t')]} \hat{\alpha}_{\nu}(t),
\end{equation}
where
\begin{eqnarray}
\Gamma_{\eta\nu}(t) & = & \int_0^{\infty} d\tau
\sum_{{\bf  k}} \bar{\lambda}^{\dag}_{{\bf  k}\eta}(t)
\sigma_{\eta}\exp\{-i\sigma_3[\omega_{{\bf
k}}-\mu-\sigma_3 E_{\nu}-\Delta_{em}-i\sigma_3 \epsilon]
\tau\}\bar{\lambda}_{{\bf  k}\nu}(t)   \nonumber \\
&=& -i\sum_{{\bf k}} \bar{\lambda}^{\dag}_{{\bf k}\eta}(t)\sigma_{\eta}
\sigma_3\frac{1}{\omega_{{\bf  k}}-\mu-\sigma_3
E_{\nu}-\Delta_{em}-i\sigma_3\epsilon}
\bar{\lambda}_{{\bf  k}\nu}(t).
\end{eqnarray}
The complex fraction should be understood as
\[ \frac{-i}{x\mp i\epsilon}=\pm\pi\delta(x)-i\frac{P}{x}, \]
where $P/x$ means the principal part of $1/x$ when integrating over $x$.

Further simplification is achieved when we notice that if the terms
$\Gamma_{\eta\nu}$ are much smaller than the energy splittings
$E_{\eta}-E_{\nu}$ between the excitation levels in the trap, then the
cross-terms with $\eta\neq \nu$ oscillate as fast
as $\sim e^{i(E_{\eta}-E_{\nu})t}$ and their contribution averages to
zero.  We then obtain a system of separate uncoupled equations for
each operator $\hat{\alpha}_{\eta}$, which is given, for non-negative
$\eta=j\geq 0$, by
\begin{equation}
\dot{\hat{\alpha}}_j= - \Gamma_{jj}(t) \hat{\alpha}_j(t)
-i\int d^3 {\bf  r} F_j({\bf  r},t) \hat{\xi}_f^{(0)}({\bf  r},t),  
\label{eq:amu0}
\end{equation}
for which the solution is
\begin{equation}
\hat{\alpha}_j(t)=\exp[-\int_0^t \Gamma_{jj}(t') dt'] \hat{\alpha}_j(0)
-i \int_0^t dt' \int d^3 {\bf  r} F_j({\bf  r},t')e^{-\int_{t'}^t
\Gamma_{jj}(t'')dt''}
\hat{\xi}_f^{(0)}({\bf  r},t').
\label{ajt}
\end{equation}
For $j=0$,
\begin{equation}
\Gamma_{00}(t)=\pi\sum_{{\bf  k}} |\lambda_{{\bf  k} 0}|^2
\delta(\omega_{{\bf  k}}-\mu-\Delta_{em}),
\end{equation}
and for $j \neq 0$,
\begin{equation}
\Gamma_{jj}=\Gamma_{j+}+\Gamma_{j-}
\label{gamma_j}
\end{equation}
where
\begin{equation}
\Gamma_{j\pm}=-i\sum_{{\bf  k}}\frac{|\lambda_{{\bf  k}
j\pm}|^2}
{\omega_{{\bf  k}}-\mu\mp E_j-\Delta_{em}\mp i\epsilon}.
\end{equation}
The imaginary part of $\Gamma_{jj}$ represents energy shifts induced by the
output coupling, while its real part $\gamma_j
\equiv {\rm Re}\Gamma_{jj}$ is given by 
\begin{equation} 
\gamma_{j\pm}\equiv {\rm Re}\Gamma_{j\pm}= \pm\sum_{{\bf
k}}|\lambda_{{\bf  k} j\pm}|^2
\delta(\omega_{{\bf  k}}-\mu\mp E_j-\Delta_{em}) 
\end{equation}
representing decay ($\gamma_{j+}>0$) or growth ($\gamma_{j-}<0$) of the
population of excited level $j$.

We now proceed to calculate the number of condensate and quasi-particle 
excitations inside the trap under the adiabatic approximation. The evolution 
of the condensate number is given straight-forwardly by
\begin{equation} 
N_0(t)=\langle \hat{\alpha}_0^2 \rangle=N_0(0)e^{-2\gamma_0 
t}. \label{N_0t} 
\end{equation}
However, for calculating $n_j(t)=\langle \hat{\alpha}_j^{\dag}(t) 
\hat{\alpha}_j(t) \rangle$
we must also consider the free term in Eq.~(\ref{ajt}), whose contribution
is proportional to the correlations of the free field operators 
$\hat{\psi}_f^{(0)}({\bf  r},t)$
\begin{equation}
\langle \hat{\psi}_f^{(0)}({\bf  r},t)(\hat{\psi}_f^{(0)})^{\dag}({\bf  
r}',t') \rangle =
\sum_{{\bf  k}}\varphi_{{\bf  k}}({\bf  r})e^{-i\omega_{{\bf
k}}(t-t')}\varphi^*_{{\bf k}}({\bf r}')= K_f({\bf r},{\bf r}',t-t').
\end{equation}
In the case of very weak coupling, where $\Gamma_{jj}$ may be assumed to
be time-independent, the contribution of this last term in
Eq.~(\ref{ajt}) to $n_j(t)$ is
\begin{eqnarray}
n^{(0)}(t) & =& \sum_{{\bf  k}} |\lambda_{{\bf  k} j-}|^2
\frac{|e^{-\gamma_j t}-e^{i(\omega_{{\bf  k}}-\bar{\omega}_{out}^{j-})t}|^2}
{(\omega_{{\bf  k}}-\bar{\omega}_{out}^{j-})^2+\gamma_j^2} \nonumber \\
\label{free-cont}
&\approx & 2\int d\omega \sum_{{\bf  k}} |\lambda_{{\bf  k}\j-}|^2
\delta(\omega-\omega_{{\bf  k}}+
\bar{\omega}_{out}^{j-}) \frac{|e^{-\gamma_j t}-e^{i\omega t}|^2}
{\omega^2+\gamma_j^2},
\end{eqnarray}
where $\bar{\omega}_{out}^{j-}=\omega_{out}^{j-}+{\rm
Im}\Gamma_{jj}$. The spectral width of the integrand is
$\Delta\omega\sim \pi/t$ for $|\gamma_j t|\ll 1$ and $\Delta\omega\sim
\gamma_j$ for $|\gamma_j t|\gg 1$. Under the conditions that led to Eq.~(\ref{eq:amu0}), one may take $\omega\approx 0$ in the
$\delta$-function and consequently identify $\sum_{{\bf k}} |\lambda_{{\bf
k}\j-}|^2 \delta(\omega_{{\bf
k}}-\bar{\omega}_{out}^{j-})=\gamma_{j-}$.  The integration over
$\omega$ may be then performed to give the final result
\begin{equation}
n_j(t)=\exp[-2\gamma_j t]n_j(0)- 2\gamma_{j-}\frac{1-e^{-2\gamma_j 
t}}{2\gamma_j}. \label{alphaj-sol}
\end{equation}
This equation is the solution of the differential equation
\begin{equation}
\frac{dn_j}{dt}=-2\gamma_{j+}n_j(t)-2\gamma_{j-}[n_j(t)+1].
\label{qp-rate}
\end{equation}
Here, the first term on the right-hand-side is responsible for an
exponential decrease in the number of excitations due to stimulated
quantum evaporation, while the second term is responsible for an
exponential increase in the number of excitations due to the process
of pair breaking, which may start even when the excited states are
initially unpopulated. This increase in the number of excitations
must, quite clearly, lead to the increase in the number of atoms in
the excited states, together with an increase in the number of output
atoms. There is, however, no process that may balance this growth in the
total number of atoms, and this implies that the growth must be compensated by a decrease in the number of condensate atoms. This is, in fact, not evident from the above equations and the problem is discussed at the end of this section.

\subsection{Solution via perturbation theory}
\label{sec:perturb}

A full solution of the linear integro-differential equations
Eq.~(\ref{eq:amu}) may be sought by perturbative iterations, taking
the magnitude of the coupling strength $\lambda$ as a perturbative
small parameter. Here we present the second-order perturbative
solutions, which are valid at short times when the population in
different excitation levels are not changed significantly from their
initial value. In this case we may also assume that the wave functions
and energies of the condensate and excitations are not changed
significantly from their initial values (i.e. $M_{\eta \nu} \approx 0$).

If we take the zeroth order solution to Eq.~(\ref{eq:amu}) to be
given by Eqs.~(\ref{a_0t}) and (\ref{a_jt}), then the second order
solution is given by
\begin{eqnarray}
\hat{\alpha}_{\eta}(t) &=&
\sum_{\nu}\left\{\delta_{\eta\nu}-
\int_0^{t} dt' \int_0^{t'} dt''  G_{\eta\nu}(t',t'')\right\}
\hat{\alpha}_{\nu}(0)  \nonumber \\
&&-i\int_0^t dt' \int d^3 {\bf  r} F_{\eta}({\bf  r},t) 
\hat{\xi}_f^{(0)}(t).
\label{sol-first}
\end{eqnarray}
Under the above assumption, we may perform the integration to obtain
\begin{eqnarray}
\hat{\alpha}_{\eta}(t) &=& \sum_{\nu}\left\{\delta_{\eta\nu}
-\sum_{{\bf  k}} \bar{\lambda}^{\dag}_{{\bf  k}\eta}
\sigma_{\eta}\bar{D}^{(2)}_{{\bf  k}\eta\nu}(t)\bar{\lambda}_{{\bf  k}\nu}
\right\} \hat{\alpha}_{\nu}(0)  \nonumber \\
&&-i\sum_{{\bf  k}}\bar{\lambda}^{\dag}_{{\bf  k}\eta}\sigma_{\eta}\sigma_3
\bar{D}_{{\bf  k}\eta}(t)
\left ( \begin{array}{c} \hat{c}_0^{\dag} \hat{b}_{{\bf  k}} \\
\hat{b}_{{\bf  k}}^{\dag} \hat{c}_0 \end{array} \right ).
\label{sol-first1}
\end{eqnarray}
Here
\begin{equation}
\bar{D}_{{\bf  k}\eta}=\left ( \begin{array}{cc} D_{{\bf  k}\eta} & 0 \\ 0 &
D^*_{{\bf  k},-\eta} \end{array} \right ),
\end{equation}
where the functions $D_{{\bf k}\eta}$ are defined in Eq.~(\ref{D-def})
and
\begin{eqnarray}
\bar{D}^{(2)}_{{\bf  k}\eta\nu}(t) &=&
\left\{\begin{array}{cc} \frac{i}{E_{\eta}-E_{\nu}}
\left[D_{{\bf  k}\eta}(t)-e^{i(E_{\eta}-E_{\nu}) t}D_{{\bf  k}\nu}(t)
\right] & \eta\neq \nu \\
\frac{1-i\{\sigma_3[\omega_{{\bf  k}}-\Delta_{em}-\mu]-E_{\eta}\} t
-e^{-i\{\sigma_3[\omega_{{\bf  k}}-\Delta_{em}-\mu]-E_{\eta} \}t}}
{\{\sigma_3[\omega_{{\bf  k}}-\Delta_{em}-\mu]-E_{\eta}\}^2} & \eta=\nu .
\end{array} \right.
\label{D2_def}
\end{eqnarray}

By using the identity
\begin{equation}
|D_{{\bf  k}\eta}(t)|^2=2{\rm Re}\{D^{(2)}_{{\bf  k}\eta\eta}(t)\},
\label{D2=2R}
\end{equation}
[see Eq.~(\ref{D2})] we obtain the following expression for the number of
condensate atoms in the trap
\begin{equation}
N_0(t) = \langle \hat{\alpha}^2_0(t) \rangle = N_0(0)\left[1
-\sum_{{\bf  k}} |\lambda_{{\bf  k} 0}|^2 |\bar{D}_{{\bf  k} 
0}(t)|^2\right],
\label{Npert0}
\end{equation}
and for the population of the excited levels we obtain
\begin{eqnarray}
n_j(t)&=& n_j(0)\left\{1
-\sum_{{\bf  k}}\left[|\lambda_{{\bf  k} j+}|^2|D_{{\bf  k} j+}(t)|^2
-|\lambda_{{\bf  k} j-}|^2|D_{{\bf  k} j-}(t)|^2\right]\right\}
\nonumber \\
&&+\sum_{{\bf  k}}|\lambda_{{\bf  k} j-}|^2 |D_{{\bf  k} j-}(t)|^2.
\label{Npertj}
\end{eqnarray}

Comparison of Eqs.~(\ref{Npert0}),~(\ref{Npertj}) with the equivalent
expressions for the number of output atoms in
Eqs.~(\ref{spect}),~(\ref{spect1}) shows that exactly one condensate
particle is taken out of the trap per each output atom generated by
the coherent output process, while one excitation (quasi-particle) is
taken from the trap per each output atom generated by the stimulated
quantum evaporation, and one excitation (quasi-particle) is
created per each atom that leaves the trap through the pair-breaking
process.

From Eq.~(\ref{sol-first1}) it is straightforward to compute the
correlations $\langle \hat{\alpha}_{\eta}^{\dag}(t) \hat{\alpha}_{\nu}(t) 
\rangle$
between the condensate and the excited levels and between the different
levels in the trap. However, it may be shown that only
diagonal terms $\eta=\nu$ are growing in magnitude with time, while the
off-diagonal correlations remain small even after a long time and
represent the effects of mixing between different levels induced by the coupling
interaction.

\subsection{Number of particles and energy}

The above treatment of the evolution of the system of a trapped Bose gas has used a
formalism which conserves the total number of particles
in the system.  However, Eqs.~(\ref{N_0t}),~(\ref{Npert0}) show that
the change in the number of condensate atoms in the system is
independent of the changes in the number of quasi-particles in the
trap. This leads to an apparent violation of number
conservation; this violation is most pronounced in the process of
pair-breaking in which output atoms are created together with
quasi-particles in the trap. The problem arises because we ignored the
off-diagonal part in the Hamiltonian which is also responsible for the 
changes in the number of condensate atoms, i.e.
\begin{equation}
\hat{{\cal H}}_{{\rm off-diag}}=
U_0 \int d^3 {\bf r} \left(\psi_0^*({\bf r})\right)^2 \hat{a}_0^{\dag}
\hat{a}_0^{\dag}\hat{\psi}_{nc}({\bf r})\hat{\psi}_{nc}({\bf r})+H.c.
\end{equation}
This part of the Hamiltonian, which is responsible for the generation of
quantum entanglement between the condensate and the excited states,
is washed-out in any mean-field treatment such as the HFB-Popov
treatment used here.  In the mean-field theory the
time-evolution of the condensate operator $\hat{a}_0$ in the steady-state is
simply given by Eq.~(\ref{a_0t}), and this leads to the apparent
violation of number conservation when the number of quasi-particles in
the system is changing. A rigorous theory which corrects this fault is
beyond the scope of this paper. Such a theory is in principle
straightforward, but technically a little complex: we have to
incorporate the anomalous average $\langle \hat{\psi}_{nc}({\bf
r}) \hat{\psi}_{nc}({\bf r}) \rangle$ into the calculation of the condensate
wave function and show how this anomalous average acquires an imaginary
part in the presence of output coupling of excited states. This, from 
another viewpoint, represents the change in the effective $T$-matrix for the 
interaction
potentials in the presence of decay.  Here, we will incorporate
number-conservation by requiring that the number of condensate atoms
$N_0(t)$ is to be determined from the conservation of the total number of
particles. If the evolution is adiabatic then some time after the switching-on of the coupling interaction the mixing between different quasi-particle levels may be neglected. The total number of atoms in the trap is then given by
\begin{equation}
N_t(t)=N_0(t)+\sum_j \left\{ n_j(t)\int d^3 {\bf
r}[|u_j({\bf r})|^2+|v_j({\bf r})|^2] +\int d^3 {\bf r} |v_j({\bf
r})|^2 \right\}. \label{Ntrap}
\end{equation}
On the other hand, we must require
\begin{equation}
N_t(t)=N_t(0)-N_{out}(t).
\end{equation}
If we compare Eqs.~(\ref{spect}),~(\ref{spect1}) with
Eqs.~(\ref{Npert0}),~(\ref{Npertj}) we see that in the process
of stimulated quantum evaporation ($\eta=j+$) the number of
quasi-particles in the trap {\em decreases} in the same rate as the
number of output atoms increases, while in the pair-breaking process
($\eta=j-$) the number of  quasi-particles in the trap {\it increases}
in the same rate as the number of output atoms {\it increases}.
In other words, in the stimulated quantum evaporation process one
thermal quasi-particle is transformed into a real output atom, while
in the pair-breaking process one quasi-particle is generated per each
output atom that leaves the trap. From inspection of
Eq.~(\ref{Ntrap}), this implies that for each atom that leaves the trap in
a stimulated quantum evaporation(SQE) process, the number of {\em particles}
associated with the
{\em quasi-particle} $j$ in the trap decreases as
\begin{equation}
\delta N^{SQE}_j = -\int d^3 {\bf  r}[|u_j({\bf
r})|^2+|v_j({\bf  r})|^2]=-1-2\int d^3 {\bf  r} |v_j({\bf  r})|^2.
\end{equation}
This must be compensated by an {\it increase} in the condensate atom
number by
\begin{equation}
\delta N^{SQE}_0= +2\int d^3 {\bf  r} |v_j({\bf  r})|^2.
\end{equation}
On the other hand, in the pair-breaking (PB) process, the number of
particles associated with the quasi-particle $j$ in the trap {\it
increases} by
\begin{equation} \delta N^{PB}_j=1+2\int d^3 {\bf  r} |v_j({\bf  r})|^2 .
\end{equation}
This must be compensated by a {\it decrease} in the condensate atom
number by
\begin{equation}
\delta N^{PB}_0=-2\int d^3 {\bf  r} |u_j({\bf
r})|^2= -2 \left (1 + \int d^3 {\bf r} |v_j({\bf r})|^2 \right ).
\end{equation}
These considerations lead us to corrections to Eq.~(\ref{N_0t})
which contains only the changes in the condensate particles
originating from direct output from the condensate component of
the Bose gas. The rate equation for the condensate atoms is now
\begin{equation}
\frac{d N_0}{dt}= -2\gamma_0 N_0-2\sum_j \left \{ \int d^3 {\bf  r}
|v_j({\bf  r})|^2|
\frac{dn_j^{SQE}}{dt}+\int d^3 {\bf  r} |u_j({\bf  
r})|^2\frac{dn_j^{PB}}{dt} \right \},
\label{N0-rate}
\end{equation}
where $dn_j^{SQE}/dt$ and $dn_j^{PB}/dt$ are the first and second terms
on the right-hand side of Eq.~(\ref{qp-rate}).
The solution of Eq.~(\ref{N0-rate}) should now replace the previous solution
for $N_0(t)$ in Eq.~(\ref{N_0t}).

\begin{figure}[t] \begin{center}
\psfig{width=12cm,file=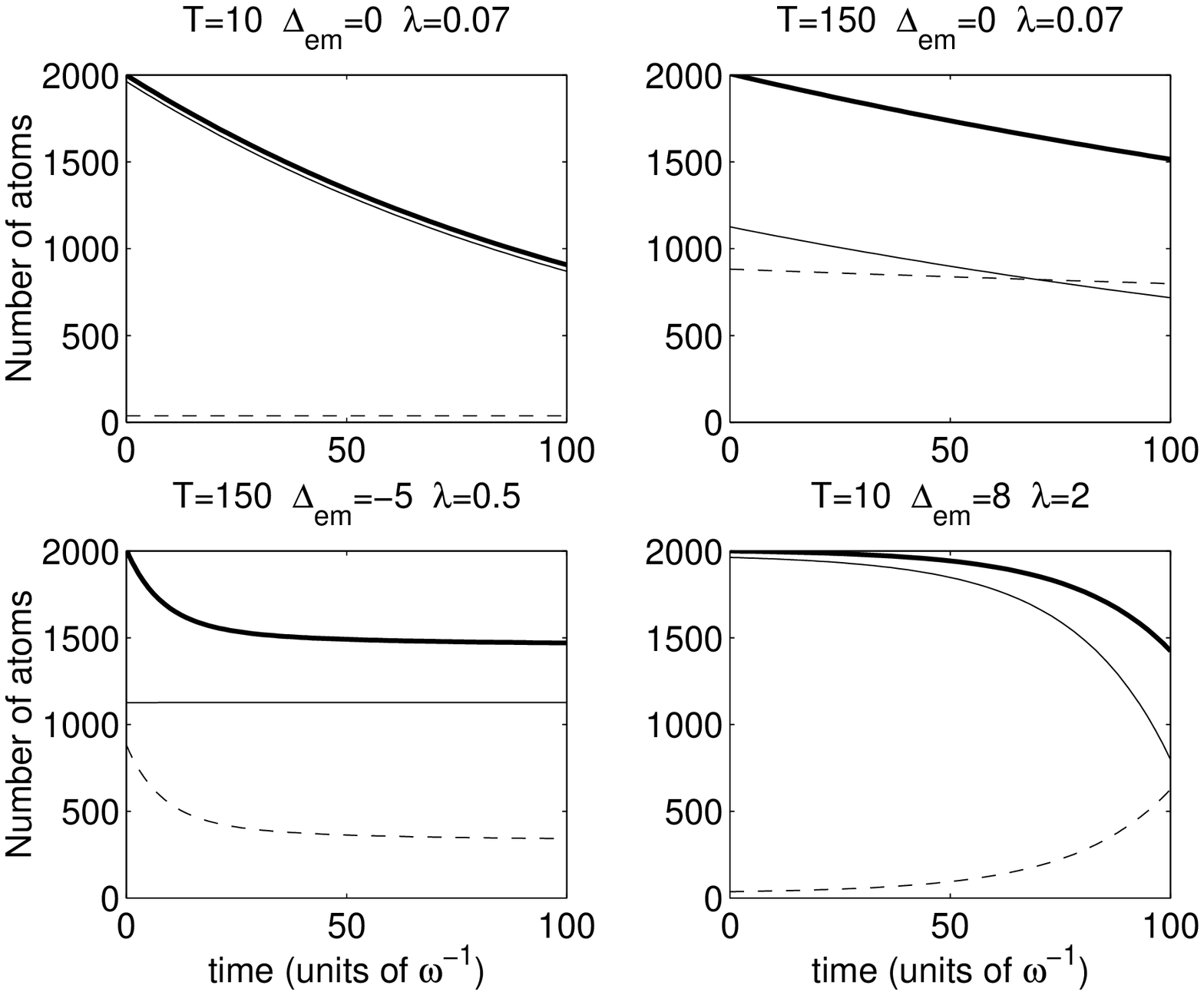}
\end{center}
\caption{\protect \footnotesize
Evolution of the condensate (solid thin line) and excited (dashed line)
atomic populations in the trap for few temperatures and coupling parameters.
The bold solid line shows the total number of atoms in the trap as a 
function
of time. In~(a) ($T=10\hbar\omega/k$) and~(b) ($T=150\hbar\omega/k$) the
output
is dominantly from the condensate. In~(c) ($\Delta_{em}=-5\omega$) 
stimulated
quantum evaporation from the higher excited levels is dominant and the
remaining population in the trap is mainly the condensate and the lower
excited levels. In~(d) ($\Delta_{em}=8\omega$) the population of the lower
excited levels increases due to pair-breaking, while the condensate
depopulates.}
\label{fig:nt}
\end{figure}

The plots of the time evolution of the trapped condensate and
non-condensate populations for few temperatures and coupling
parameters are given in Fig.~\ref{fig:nt}. These plots are solutions
of the differential equations~(\ref{qp-rate}) and~(\ref{N0-rate}).
When $\Delta_{em}=0$ (Fig.~\ref{fig:nt}a,b) the condensate part
decreases while the thermal part does not change significantly. When
$\Delta_{em}=-5\omega$ (Fig.~\ref{fig:nt}c) conservation of energy
only permits transitions from upper excited states to the output
level, and the population of the condensate and the lower excited states
thus remains unchanged. The upper excited states, in this case,  are 
found to depopulate completely as to be expected.
When $\Delta_{em}=8\omega$ (Fig.~\ref{fig:nt}d) the
thermal population grows significantly due to transitions of unpaired
atoms from the condensate into the excited states. However, the energy
distribution in the lower excited states is a highly non-equilibrium
distribution and dissipation and thermalization effects that have not
been taken into account in this paper should play a major role. The
short time limit i.e. $0 \leq t < 10\omega^{-1}$ behaviour is clearly not
accurately described in these plots but it can be calculated by using
the low-order perturbative expansions of Sec.~\ref{sec:perturb}.

Changes in the total energy in the trap may be caused either by the
transfer of atoms out of the trap or by the changes in the chemical
potential $\mu$ and energies $E_j$ of the excitations. The second kind
of process is beyond the scope of this paper, since we have neglected
changes in $\mu$ and $E_j$ and put $M_{\eta\nu}=0$ in
Eq.~(\ref{eq:amu}).  As for the first kind of process, an energy
quantum of $\delta E=\mu$ leaves the trap for each condensate atom
that leaves the trap (consequently the energy in the trap is {\em reduced} by $\mu$), the energy changes by $\delta
E_j^{SQE}=-\mu-E_j$ for each atom that leaves the trap by the
stimulated quantum evaporation process, and finally $\delta
E_j^{PB}=-\mu+E_j$ for each atom that leaves the trap through the
pair-breaking process. Therefore we have a relatively simple result for the 
rate of change in energy:
\begin{equation}
\frac{dE_t}{dt}= \mu\frac{dN_t}{dt}+\sum_j E_j \frac{dn_j}{dt}.
\end{equation}

\section{Summary and Discussion}
\label{sec:discuss}

In this paper we have set up a general theory of weak output coupling
from a trapped Bose-Einstein gas at finite temperatures. The
formalism developed here is suitable for the discussion of both
Radio-frequency or stimulated Raman output couplers. It has enabled us
to gain much information on the basic features that we expect in real
experiments: the time-dependence of the output beam, the effects of
excitations in the trapped Bose gas and the pairing of particles.
Predictions for specific systems can also be based on our theory.

For the time-dependence of the output beam, we have shown that the
output beam is a mixture of components from different origins in the
trap.  The output condensate ($\eta=0$) is the coherent part of the
beam, while each excited level $j$ in the trap contributes two partial
waves: one originating from the process of stimulated quantum
evaporation ($\eta=j+$), where a quasi-particle (excitation) in the
trap transforms into a real output atom, and the other originating
from the pair-breaking process ($\eta=j-$), where two correlated atoms
in the trap transform into a quasi-particle in the trap and a real
output atom. We have shown that a steady monochromatic wave from each
component is formed after a time comparable to the inverse of
the bandwidth of the corresponding matrix element $\lambda_{{\bf
k}\eta}$ as a function of $\omega_{\bf k}$.  We have also analyzed the
oscillatory behaviour of the output rate at short times and showed the
existence of non-propagating bound states in the untrapped level that
are formed near the trap as a result of the mixing induced by the
output coupler.

As for the evolution of the Bose gas in the trap during the process of output
coupling, we have shown that for the case of weak coupling an adiabatic
approximation may be made, which enables calculation of the
composition of the Bose gas inside the trap in terms of the adiabatic
basis of condensate and excitations.  We have shown that exponential
decay of the excitations is expected when the stimulated quantum
evaporation process is dominant, while an exponential growth of the
number of excitations is expected when the pair-breaking process is
dominant. We have shown that the number of trapped condensate atoms
increases in each event of stimulated quantum evaporation, while it
decreases by more than 2 atoms per each event of
pair-breaking. However, we stress that a more elaborate
number-conserving theory of time-dependent evolution of the Bose gas
in an open system than that considered here is needed.

The coherence of the output beam was shown to depend on parameters
under experimental control such as the detuning of the laser. We note
that the coherence of the output atoms also tells us about the coherence
properties of atoms inside the trap; the coherence of the trapped Bose
gas is expected to be altered as a direct consequence of the output
coupling. In simple terms, when the output atoms are mainly those of
condensates we expect the coherence of the internal atoms to drop, if
only because the amount of coherent condensate fraction decreases. The
coherence of trapped atoms, although interesting theoretically, is not
experimentally verifiable.

Apart from designing an atomic laser with well-controlled beam properties, 
we saw  that the measured output properties may be an
excellent tool in investigating the nature of trapped Bose gases at finite 
temperatures. The properties of the output beam may be
a probe to the temperature of the trapped Bose gas as well as the internal
structure of the ground state and the excitations. The present
treatment may be extended to cope with other possible
configurations that are likely to appear in the future such as a trap with
multi-component condensates and Bose gases with negative scattering
lengths.

Finally we note that the pair breaking process, and indeed the output 
coupling of the
condensate in general, provides an experimentally feasible method to
study quantum entanglement in a macroscopic system. The quantum theory
of entanglement is currently under intense study owing to its
relevance to quantum computation; so far it has rarely been studied
in the context of BEC.

\acknowledgments

S.C. acknowledges UK CVCP for support.
Y.J. acknowledges support from the Foreign and Commonwealth
Office by the British Council, from the Royal Society of London and from
EC-TMR grants. This work was supported in part by the
U.K. EPSRC (K.B.).

\appendix
\section{Expression for $\lambda({\bf  r},t)$ -- stimulated Raman scheme}
\label{app:Raman}

We derive here the expression for the effective coupling function
$\lambda({\bf  r},t)$ in Eq.~(\ref{lambda}) for the stimulated Raman
transition
coupling scheme. A detailed analysis of a Raman coupling process from a
condensate in the mean-field approach can be found in Ref.\cite{Edwards99}.

We consider a single atom which can be found either in the trapped level
$| t \rangle$ or in the free level $| f \rangle$. A pair of  laser beams 
with
spatial and temporal amplitudes $E_{tL}({\bf  r},t)$ and $E_{fL}({\bf  
r},t)$
are
responsible for non-resonant transitions from $| t \rangle$ and $| f 
\rangle$
to a
high energy level $| i \rangle$. The Hamiltonian is then given by
\begin{equation}
H = \sum_{j=t,f,i} [\hbar\omega_j+H_j^{(0)}] | j \rangle\langle j |
+\frac{1}{2}\sum_{j=t,f}[\mu_{ji} E_{jL}({\bf r},t) | i \rangle \langle j
|+h.c.] \end{equation}
Here $\hbar\omega_j$ are the internal  energies of the levels $| j \rangle$,
and
\begin{equation}
H^{(0)}_j= -\frac{\hbar^2 \nabla^2}{2m}+ V_j({\bf r}),
\end{equation}
where $V_j({\bf r})$ are effective external potentials acting on the 
different
atomic levels, and $\mu_{j3}$ are the dipole moments for the transition
$| j \rangle\rightarrow  | i \rangle$. The amplitudes of the laser fields 
are
assumed
to have the form $E_{jL}({\bf r},t)={\cal E}_j({\bf r},t)
e^{i({\bf k}_{jL}\cdot {\bf r} - \omega_{jL} t)}$ where
the envelopes ${\cal E}_{jL}$ are slowly varying with respect to the
exponential term. The wave function describing the atom has the form
$\sum_{j=t,f,i} \psi_j({\bf  r},t)e^{-i\omega_j t}| j \rangle$. The 
equations
of motion
for the three amplitudes are:
\begin{eqnarray}
i\hbar\dot{\psi_t} &=& H_t^{(0)}\psi_t-\mu^*_{ti}E^*_{tL} \psi_i
e^{-i(\omega_i-\omega_t)t} \label{psi1} \\
i\hbar\dot{\psi_f}&=&H_f^{(0)}\psi_f-\mu^*_{fi}E^*_{fL} \psi_i
e^{-i(\omega_i-\omega_f)t} \label{psi2} \\
i\hbar\dot{\psi_i}&=&H_i^{(0)}\psi_i-\mu_{ti}E_{tL} \psi_t
e^{-i(\omega_t-\omega_i)t}-\mu_{fL} E_{fL}\psi_f e^{-i(\omega_f-\omega_i)t}.
\end{eqnarray}
The solution of the equation for $\psi_i$ as a function of the two other
amplitudes can be written as
\begin{eqnarray}
\psi_i({\bf r},t) &=& \psi_i^{(0)}({\bf r},t)+i\int_0^t dt' \int d^3 {\bf
r}'
K_i({\bf r},{\bf r}',t-t')\sum_{j=t,f} e^{-i(\omega_j-\omega_i)t'} \nonumber
\\
&&\times \Omega_j({\bf r}',t')\psi_j({\bf r}',t')
e^{i({\bf k}_{jL}\cdot {\bf r}'-\omega_{jL}t')}. \label{psi3}
\end{eqnarray}
Here $\psi_i^{(0)}$ is the solution of the Schr\"odinger equation $i\hbar
\partial\psi_i/\partial t=H_i^{(0)}\psi_i$ with the initial condition
$\psi^{(0)}(0)=
\psi_i(0)=0$, under the assumption that level $| i \rangle$ is initially
unpopulated.
$\Omega_j({\bf r},t)=\mu_{ji}{\cal E}_{jL}({\bf r},t)/2\hbar$ are the slowly
varying Rabi frequencies and $K_i$ is the propagator for the evolution
of the level $| i \rangle$, which can be expanded in
terms of the energy eigenfunctions of $H_i^{(0)}$ in a similar way to the
expansion in Eq.~(\ref{Gfk}).

The crucial step now is to notice that the main time-dependence in the
time-integral in Eq.~(\ref{psi3}) comes from the terms
$e^{-i(\omega_j-\omega_i+\omega_{jL})t}$, whose frequency of oscillation is
assumed
to be in the optical range, while the other terms, which  correspond to
atomic centre-of-mass motion are oscillating in frequencies below the
radio-frequency range. Assuming that the switching time of the coupling is
much longer than the short period of oscillation of the fast terms, we 
expect
the contribution to the integral in $t'$ to come only from a short time
interval around the end-point $t$. We then take $t'=t$ in the slow terms.
As a result, we
have $K_i({\bf  r},{\bf  r}',t-t')\approx K_i({\bf  r},{\bf
r}',0)=\delta({\bf  r}-{\bf  r}')$ and
$\psi_j({\bf  r}',t')\approx \psi_j({\bf  r},t)$, and obtain
\begin{equation}
\psi_i({\bf r},t) = -\sum_{j=t,f}
\frac{e^{-i(\omega_j-\omega_i+\omega_{jL})t}-1}{\omega_j-\omega_i+\omega_L}
\Omega_j({\bf r},t)\psi_j({\bf r},t) e^{i{\bf k}_{jL}\cdot {\bf r}}.
\end{equation}
When this is substituted in Eqs.~(\ref{psi1}),~(\ref{psi2}), and the rapidly
oscillating terms are dropped, we obtain
\begin{eqnarray}
\dot{\psi_t} &=& -\frac{i}{\hbar}H_t^{(0)}\psi_t+i\lambda_{tt}({\bf
r},t)\psi_t({\bf  r},t)
+i\lambda_{tf}({\bf  r},t)\psi_f({\bf  r},t) \\
\dot{\psi_f}&=& -\frac{i}{\hbar} H_f^{(0)}\psi_f+i\lambda_{ff}({\bf  r},t)
\psi_f({\bf  r},t) +i\lambda_{ft}({\bf  r},t)\psi_t({\bf  r},t),
\end{eqnarray}
where
\[ \lambda_{jj'}({\bf  r},t)=\frac{\Omega^*_j({\bf  r},t)\Omega_{j'}
({\bf r},t)}{\omega_{j'}-\omega_i+\omega_{j'L}}e^{-i(\omega_{j'L}
+\omega_{j'}-\omega_{jL}-\omega_j)t} e^{i({\bf k}_{j'L}-{\bf
k}_{jL})\cdot{\bf r}}. \]
The form of $\lambda({\bf r}.t)$ in
Eq.~(\ref{lambda}) is achieved by assuming
$\omega_t-\omega_i+\omega_{tL}\approx
\omega_f-\omega_i+\omega_{fL}\equiv
\Delta_i$ and then
noticing that $\lambda_{tf}=\lambda_{ft}^*$. We have also neglected
the diagonal terms $\lambda_{tt},\lambda_{ff}$, which are responsible
for an additional effective potential acting on the levels $| t
\rangle$ and $| f \rangle$, under the assumption that they are small
compared to the other potentials $V_t({\bf r})$ and $V_f({\bf r})$
near the trap. This assumption is justified in the adiabatic case
discussed in this paper, where the coupling is assumed to be weak and
slow.

\vspace{10mm}


\vspace{10mm}

\newpage

in


\begin{thebibliography}{99}


\bibitem{MITOutput97} M.-O. Mewes, M. R. Andrews, D. M. Kurn, D. S. Durfee,
C. G. Townsend, and W. Ketterle Phys. Rev. Lett. {\bf 78}, 582 (1997)

\bibitem{MITfringes97} M. R. Andrews, C. G. Townsend, H.-J. Miesner,  D. S.
Durfee,  D. M. Kurn, and W. Ketterle Science {\bf 275}, 637 (1997)

\bibitem{NewZealand99} J. L. Martin, C. R. McKenzie, N. R. Thomas,
J. C. Sharpe, D. M. Warrington, P. J. Manson, W. J. Sandle, and A. C. 
Wilson,
J. Phys. B {\bf 32}, 3065 (1999).

\bibitem{NIST99} E. W. Hagley, L. Deng, M. Kozuma, J. Wen, K. Helmerson, S. 
L.
Rolston and W. D. Phillips  Science {\bf 283}, 1706 (1999)

\bibitem{Munich99} I. Bloch, T. W. H\"{a}nsch and T. Esslinger Phys. Rev.
Lett. {\bf 82}, 3008 (1999)

\bibitem{Hop97} J. J. Hope Phys. Rev. A {\bf 55}, 2531 (1997)

\bibitem{MoySav97} G. M. Moy and C. M. Savage Phys. Rev. A {\bf 56}, R1087
(1997)

\bibitem{JefHorBar99} J. Jeffers, P. Horak, S. M. Barnett, and P. M. Radmore
(Unpublished)

\bibitem{BalBurSco97} R. J. Ballagh, K. Burnett, and T. F. Scott Phys. Rev.
Lett. {\bf 78}, 1607 (1997)

\bibitem{Steck98} H. Steck, M. Naraschewski, and H. Wallis Phys. Rev. Lett.
{\bf 80}, 1
(1998)

\bibitem{Durham98} B. Jackson, J. F. McCann, and C. S. Adams, J. Phys.
B {\bf 31}, 4489 (1998).

\bibitem{Jack99} M. W. Jack, M. Naraschewski, M. J. Collett, and D. F. Walls
Phys. Rev. A {\bf 59}, 2692 (1999)

\bibitem{Band99} Y.B. Band, M. Trippenbach and P.S. Julienne, Phys. Rev.
{\bf A 59}, 3823 (1999).

\bibitem{Edwards99} M. Edwards, D. A. Griggs, P. L. Holman, C. W. Clark,
S. L. Rolston, and W. D. Philips, J. Phys. B {\bf 32}, 2935 (1999).


\bibitem{JapChoiBur99} Y. Japha, S. Choi, K. Burnett and Y. Band, Phys. Rev.
Lett. {\bf 82} 1079 (1999)

\bibitem{Qevap} F. Dalfovo et. al., Phys. Rev. Lett. {\bf 75}, 2510 (1995);
J. Low. Temp. Phys. {\bf 104}, 367 (1996);
A. F. G. Wyatt, Nature {\bf 391}, 56 (1998);
A. Griffin, Nature {\bf 391}, 25 (1998)

\bibitem{Wilkens-scatt} Z. Idziaszek, K. Rzazewski and M. Wilkens, J. Phys.
B {\bf 32}, L205 (1999).

\bibitem{Gri96} A. Griffin Phys. Rev. B {\bf 53}, 9341 (1996)
\label{Gri96}

\bibitem{Prou} There are recent theories which go beyond HFB from
considerations of microscopic interaction between particles eg. N. P.
Proukakis, K.
Burnett, and H. T. C. Stoof, Phys. Rev. A {\bf 57}, 1230 (1998);  N. P .
Proukakis, S.
A. Morgan, S. Choi and K. Burnett, Phys. Rev. A  {\bf 58}, 2435 (1998). 
These
mainly result in small shifts in energies of excitations. We assume these
differences are not as crucial in modelling an output coupler.

\bibitem{numconserve}  C. W. Gardiner, Phys. Rev. A {\bf 56}, 1414 (1997);
M. D. Girardeau, Phys. Rev. A {\bf 58}, 775 (1998).

\bibitem{MorChoiBur97} S. A. Morgan, S. Choi, K. Burnett and M. Edwards, 
Phys.
Rev. A {\bf 57} 3818 (1999)

\bibitem{Cohen-Tannoudji}
C. Cohen-Tannoudji, J. Dupont-Roc, and G. Grynberg, {\it Atom-photon
interactions : basic processes and applications}
(Wiley, New-York, 1992). Complement $C_{III}$.

\bibitem{Glauber} R. J. Glauber, Phys. Rev. {\bf 130}, 2529 (1963)

\bibitem{Goldstein98} E. V. Goldstein and P. Meystre, Phys. Rev. Lett.
{\bf 80 }, 5036 (1998);
E. V. Goldstein, O. Zobay, and P. Meystre, Phys. Rev. {\bf A 58}, 2373 
(1998).

\bibitem{Glauber99} M. Naraschewski and R. J. Glauber, Phys. Rev. A {\bf 
59},
4595 (1999).

\bibitem{Javanainen99} J. Javanainen, J. Ruostekoski, B. Vestergaard and
M. R. Francis, Phys. Rev. A {\bf 59}, 649 (1999).

\bibitem{DodBurEdw97} R. J. Dodd, K. Burnett, M. Edwards, and C. W,. Clark
Optics Express  {\bf 1}, 284 (1997)

\bibitem{CasDum96} Y. Castin and R. Dum, Phys. Rev. Lett. {\bf 77}, 5315
(1996); {\bf 79}, 3553 (1997);

\bibitem{CasDum98} Y. Castin and R. Dum, Phys. Rev. A {\bf 57}, 3008 (1998)

\bibitem{Damping0} D. S. Jin, M. R. Matthews, J. R. Ensher, C. E. Wieman,
and E. A. Cornell, Phys. Rev. Lett. {\bf 78}, 764 (1997).

\bibitem{Damping1} S. Giorgini, Phys. Rev. A {\bf 57}, 2949 (1998);
P. O. Fedichev, G. V. Shlyapnikov and J. T. M. Walraven, Phys. Rev.
Lett. {\bf 80}, 2269 (1998); L. P. Pitaevskii and S. Stringari,
Phys. Lett. A {\bf 235}, 398 (1997);  V. Liu, Phys. Rev. Lett.
{\bf 79}, 4056 (1997).

\bibitem{JapBand99} Y. Japha, and Y. B. Band (Unpublished)

\bibitem{BurGhrMya97} E. A. Burt, R. W. Ghrist, C. J. Myatt, M. J. Holland, 
E.
A. Cornell, C. E. Wieman, Phys. Rev. Lett  {\bf 79}, 337 (1997)


\end{thebibliography}
\end{document}